\documentclass[a4paper,11pt]{article}
\usepackage{amsmath,amssymb,theorem}
\usepackage[cm]{fullpage}
\usepackage[shortcuts]{extdash}
\usepackage[utf8]{inputenc}
\usepackage[nosort]{cite}
\usepackage{hyperref}

\let\frac\undefined

\allowdisplaybreaks

\numberwithin{equation}{section}

\def\Maketitle{{\def\newpage{}\maketitle}}

\def\eq#1$$#2$${\begin{equation#1}#2\end{equation#1}}
\long\def\subeq#1{\begin{subequations}#1\end{subequations}}
\def\Split$$#1$${\begin{split}#1\end{split}}
\def\Align#1$$#2$${\begin{align#1}#2\end{align#1}}
\def\AlignAt#1$$#2$${\begin{alignat}{#1}#2\end{alignat}}
\def\Aligned#1{\begin{aligned}#1\end{aligned}}
\def\Gather#1$$#2$${\begin{gather#1}#2\end{gather#1}}
\def\Gathered#1{\begin{gathered}#1\end{gathered}}
\def\Multline#1$$#2$${\begin{multline#1}#2\end{multline#1}}

\def\Cases#1{\begin{cases}#1\end{cases}}
\def\d{\partial}

\def\Res{\mathop{\rm Res}\limits}

\def\spanop{\mathop{\rm span}\nolimits}

\def\vac{\mathinner{\rm vac}}

\def\cA{{\mathcal A}}

\def\cF{{\mathcal F}}
\def\bcF{{\bar{\mathcal F}}}

\def\cH{{\mathcal H}}

\def\cN{{\mathcal N}}
\def\cO{{\mathcal O}}

\def\cS{{\mathcal S}}

\def\ve{\varepsilon}
\def\sh{\mathop{\rm sh}\nolimits}
\def\ch{\mathop{\rm ch}\nolimits}

\def\lcolon{\mathopen{\,:}}
\def\rcolon{\mathclose{:\,}}

\def\R{{\mathbb R}}
\def\Z{{\mathbb Z}}

\def\fr{{\mathfrak r}}
\def\fs{\mathop{\mathfrak s}\nolimits}
\def\ft{{\mathfrak t}}
\def\ftdag{{\mathfrak t}^{\smash\dagger}}
\def\fT{{\mathfrak T}}
\def\e{\mathop{\rm e}\nolimits}
\def\i{{\rm i}}

\def\tq{{\tilde q}}

\def\ttau{{\tilde\tau}}
\def\ttaudag{\ttau^{\smash\dagger}}
\def\tomega{{\tilde\omega}}

\def\tS{{\tilde S}}
\def\tSdag{\tS^{\smash\dagger}}
\def\tsigma{{\tilde\sigma}}
\def\tsigmadag{\tsigma^{\smash\dagger}}
\def\tSigma{{\tilde\Sigma}}
\def\tSigmadag{\tSigma^{\smash\dagger}}
\def\Pdag{P^{\smash\dagger}}
\def\pdag{p^{\smash\dagger}}

\def\ud{{\rm d}}
\def\uddag{{\rm d}^{\smash\dagger}}
\def\Xdag{X^{\smash\dagger}}

\def\bz{{\bar z}}

\def\bh{{\bar h}}

\def\bm{{\bar m}}

\def\lambdadag{\lambda^{\smash\dagger}}
\def\etadag{\eta^{\smash\dagger}}
\def\epsilondag{\epsilon^{\smash\dagger}}
\def\tdag{t^{\smash\dagger}}

\def\Rdag{R^{\smash\dagger}}
\def\Sdag{S^{\smash\dagger}}
\def\Sinv{S^{\smash-1}}
\def\Sdaginv{S^{\smash\dagger-1}}
\def\sigmadag{\sigma^{\smash\dagger}}
\def\taudag{\tau^{\smash\dagger}}

\def\brks#1{[#1]^{\vphantom{\ddagger}}_\omega}
\def\tbrks#1{[#1]^{\vphantom{\ddagger}}_\tomega}
\def\brcs#1{\{#1\}^{\vphantom{\ddagger}}_\omega}
\def\tbrcs#1{\{#1\}^{\vphantom{\ddagger}}_\tomega}

\makeatletter

\def\section{\@startsection{section}{1}{\z@}%
                                   {-3.5ex \@plus -1ex \@minus -.2ex}%
                                   {2.3ex \@plus.2ex}%
                                   {\normalfont\normalsize\bfseries}}
\def\subsection{\@startsection{subsection}{2}{\z@}%
                                     {-3.25ex\@plus -1ex \@minus -.2ex}%
                                     {1.5ex \@plus .2ex}%
                                     {\normalfont\normalsize\bfseries\itshape}}
\def\@seccntformat#1{\csname the#1\endcsname.~~}
\long\def\@makecaption#1#2{%
  \vskip\abovecaptionskip
  \sbox\@tempboxa{\small#1. #2}%
  \ifdim \wd\@tempboxa >0.9\hsize
  {\leftskip=0.05\hsize\rightskip=0.05\hsize\relax\small
    #1. #2\par}
  \else
    \global \@minipagefalse
    \hb@xt@\hsize{\hfil\box\@tempboxa\hfil}%
  \fi
  \vskip\belowcaptionskip}
\def\Appendix{\appendix
  \def\@seccntformat##1{Appendix~\csname the##1\endcsname.~~}}

\let\over\@@over
\let\atop\@@atop
\let\above\@@above
\let\overwithdelims\@@overwithdelims
\let\atopwithdelims\@@atopwithdelims
\let\abovewithdelims\@@abovewithdelims

\makeatother

\long\def\?#1{{\par\medskip\hrule\smallskip\noindent
{\bf What is missing:} #1\smallskip\hrule\medskip\par}}

\begin{document}

\title{The complex sinh\-/Gordon model:\\
	form factors of descendant operators and\\
 current\--current perturbations}
\author{Michael Lashkevich and Yaroslav Pugai,\\[\medskipamount]
	\parbox[t]{0.8\textwidth}{\normalsize\it\raggedright
		Landau Institute for Theoretical Physics, 142432 Chernogolovka, Russia\medspace%
		\footnote{Mailing address.}
		\\
		Moscow Institute of Physics and Technology, 141707 Dolgoprudny, Russia\\
		Kharkevich Institute for Information Transmission Problems, 19 Bolshoy Karetny per., 127994 Moscow, Russia
		\\
		E-Mail. ML: lashkevi@landau.ac.ru, YP: slava@itp.ac.ru}
}
\date{}

\Maketitle

\begin{abstract}
We study quasilocal operators in the quantum complex sinh\-/Gordon theory in the form factor approach. The free field procedure for descendant operators is developed by introducing the algebra of screening currents and related algebraic objects. We work out null vector equations in the space of operators. Further we apply the proposed algebraic structures to constructing form factors of the conserved currents $T_k$ and $\Theta_k$. We propose also form factors of current\--current operators of the form $T_kT_{-l}$. Explicit computations of the four\-/particle form factors allow us to verify the recent conjecture of Smirnov and Zamolodchikov about the structure of the exact scattering matrix of an integrable theory perturbed by a combination of irrelevant operators. Our calculations confirm that such perturbations of the complex sinh\-/Gordon model and of the $\Z_N$ symmetric Ising models result in extra CDD factors in the $S$ matrix.
\end{abstract}

\section{Introduction and results}

The complex sinh\-/Gordon model is an interesting integrable massive model, which has been studied extensively at both, classical \cite{Pohlmeyer:1975nb,Lund:1976ze} and quantum \cite{deVega:1981ka,deVega:1982sh,Dorey:1994mg,Fateev:1995ht,Fateev:2001mj,Fateev:2017mug} levels. In the quantum case the exact $S$ matrix has been found, which made it possible to develop the bootstrap form factor program.

\subsection{Sigma model description}

We will not use directly the Lagrangian description in this work. However, let us recall it for completeness and fixing notation. The complex sinh\-/Gordon model can be described as the theory of a complex bosonic field $\chi=\chi_1+\i\chi_2$, $\bar\chi=\chi_1-\i\chi_2$ with the action
\eq$$
\cS[\chi,\bar\chi]=\int {d^2x\over4\pi}\,\left({\d_\mu\chi\,\d^\mu\bar\chi\over1+g_0\chi\bar\chi}-M_0^2\chi\bar\chi\right),
\label{CShG-action}
$$
Initially, the model was introduced for negative values of a bare coupling constant $g_0$ by Pohlmeyer\cite{Pohlmeyer:1975nb} and Lund and Regge\cite{Lund:1976ze}. In this regime the theory has a rich spectrum of solitons and their bound states. In what follows we basically consider the model in the regime $g_0>0$. The spectrum in this case consists of a single particle\-/antiparticle pair~\cite{Fateev:1995ht}. In the semiclassical limit\cite{deVega:1981ka,deVega:1982sh} for $g_0<0$ the coupling constant is renormalized as
\eq$$
g={g_0\over1+g_0}.
\label{g-renorm}
$$
As for the mass $M$ of the particles it is only equal to $M_0$ for small values of the coupling constant $g$. For general coupling constants we have
\eq$$
M\sim M_0^{1/(1-2g)}.
\label{M-M0-rel}
$$
For $M_0=0$ the action describes a conformal model\cite{Witten:1991yr}. The perturbation operator $\chi\bar\chi$ has the conformal dimension~$2g/(2g-1)$.

The sigma\-/model type action (\ref{CShG-action}) is convenient for perturbative calculation of the $S$ matrix of the non\-/topological particles, but when studying the operator contents of the theory we encounter difficulties. The perturbation constant is $g_0$ is dimensionless, and a complete exact solution of the conformal field theory, which describes the model at small distances, is missing. Thus, the operators of the theory cannot be generally expressed in terms of the basic field~$\chi$.

\subsection{Fateev's action and the operator contents}

Fateev\cite{Fateev:1995ht,Fateev:2001mj} proposed a dual description of the theory~(\ref{CShG-action}) in terms of two neutral scalar fields $\vartheta(x),\varphi(x)$ with the action
\eq$$
\cS_{\rm dual}[\vartheta,\varphi]=\int d^2x\,\left({(\d_\mu\vartheta)^2+(\d_\mu\varphi)^2\over8\pi}
+M\e^{\beta\varphi}\cos\alpha\vartheta-CM^{2+4\beta^2}\e^{-2\beta\varphi}\right).
\label{dual-action}
$$
The coupling constants $\alpha$ and $\beta$ are not independent and related to the coupling constant $g$ of the sigma\-/model action according to the equation
\eq$$
\alpha^2=\beta^2+1={1\over2g}.
\label{alpha-beta-def}
$$
The constant $M$ is the physical mass subject to a particular choice of the constant~$C$.

We will consider the exponential operators
\eq$$
\phi_{m\bm}^{\kappa}(x)
=\exp\left(-{\kappa\over2\beta}\varphi(x)-{\i m\over2\alpha}(\vartheta(x)+\tilde\vartheta(x))
+{\i\bm\over2\alpha}(\vartheta(x)-\tilde\vartheta(x))\right)
\label{Phi-exp-def}
$$
and their descendants. Here $\kappa$, $m$, $\bm$ are real numbers. The symbol $\tilde\vartheta(x)$ stands for the dual field defined by the equation 
$$
\d^\mu\tilde\vartheta=\epsilon^{\mu\nu}\d_\nu\vartheta.
$$
The complex sinh\-/Gordon basic fields $\chi,\bar\chi$ and the operator $\chi\bar\chi$ coincide up to normalization constants with the following exponential operators
\eq$$
\Aligned{
\chi(x)
&\sim\phi^1_{-1,-1}(x),
\\
\bar\chi(x)
&\sim\phi^1_{1,1}(x),
\\
\chi\bar\chi(x)
&\sim\phi^2_{0,0}(x).
}\label{chi-barchi-V}
$$
There is a $U(1)$ charge $Q$ associated to the complex boson~$\chi,\bar\chi$ corresponding to the physical particle and antiparticle. Assuming $Q(\chi)=-1$ (since $\chi$ annihilates a particle) and $Q(\bar\chi)=1$ (since $\bar\chi$ annihilates an antiparticle), we see that the charge of the operator $\phi_{m\bar{m}}^{\kappa}$ is
\eq$$
Q(\phi_{m\bar{m}}^{\kappa})={m+\bar{m}\over2}\in\Z.
\label{Qcharge-Phi}
$$
This relation will provide a selection rule for matrix elements.

The exponential operators contain, in general, both the field $\vartheta$ and its dual~$\tilde\vartheta$. For this reason the operators are not mutually local. Recall that two operators $\cO_1,\cO_2$ are mutually quasilocal, if they have the following property. Consider any correlation function $\langle\cO_1(x_1)\cO_2(x_2)\cdots\rangle$. The move of the point $x_1$ around $x_2$ in the counter\-/clockwise direction along a closed contour results in the multiplication of the correlation function by a phase factor $\e^{2\pi\i\gamma}$. The number $\gamma=\gamma(\cO_1,\cO_2)$ is called the mutual locality exponent of the two operators. For the exponential operators $\phi_{m\bar{m}}^{\kappa}(x)$ the mutual locality exponents are
\eq$$
\gamma(\phi_{m\bar{m}}^\kappa,\phi_{m'\bar{m}'}^{\kappa'})={g(mm' - \bar{m}\bar{m}')\over2}.
\label{VV-locality}
$$
In particular, for example, the energy operators $\phi^{2k}_{00}$ are mutually local with the spin and disorder operators. The operator $\phi_{m\bar{m}}^\kappa$ is local with respect to the basic bosons of the initial Lagrangian~(\ref{CShG-action}) for $g(\bar{m}-m)/2\in\Z$ only.

We study the structure of the space $\cH^\kappa_{m\bm}$ of operators obtained from the exponential operators $\phi^\kappa_{m\bm}$ by attaching space\-/time derivatives of the fields $\varphi,\vartheta$. It is convenient to parameterize the space\-/time by the light cone coordinates $x^\pm=x^1\pm x^0$ with the corresponding derivatives~$\d_\pm$. Then this space reads
\eq$$
\cH^\kappa_{m\bm}
=\spanop\{(\d_-^{k_1}\vartheta\cdots\d_-^{l_1}\varphi\cdots
\d_+^{\bar k_1}\vartheta\cdots\d_+^{\bar l_1}\varphi\cdots)\phi_{m\bar{m}}^\kappa(x)\}.
\label{descendant-def}
$$
We call a basic operator a level $(k,\bar k)$ descendant of the exponential operator $\phi_{m\bar{m}}^\kappa$, if
\eq$$
\Aligned{
&k=\sum k_i+\sum l_i,\\ 
&\bar k=\sum\bar k_i+\sum\bar l_i
}\label{level-def}
$$ 
The Lorentz spin $s$ of a descendant operator and its ultraviolet scaling dimension $d$ are given by the expressions
\eq$$\Aligned{
s
&=s_{m\bm}+k-\bar k,
\qquad
&s_{m\bm}
&=g{m^2-\bar{m}^2\over 4},
\\
d
&=d^\kappa_{m\bm}+k+\bar k,
\qquad
&d^\kappa_{m\bm}
&={m^2+\bar{m}^2\over8\alpha^2}-{\kappa^2\over4\beta^2}.
\label{sd-descendant}
}
$$
Here $s_{m\bm}$ and $d^\kappa_{m\bm}$ are the spin and the dimension of the exponent, while $\Delta k=k-\bar k$ and $L=k+\bar k$ are contributions of derivative of the fields~$\varphi,\vartheta$. We shall call $\Delta k$ the `external' spin and $L$ the total level.

The model (\ref{dual-action}) can also be considered as a perturbation of the sine\-/Liouville theory by the operator~$\e^{2\beta\varphi}$. The conformal dimensions of the operators in the sine\-/Liouville theory differs from those given by~(\ref{sd-descendant}) by an extra term $-2\kappa/4\beta^2$. The sine\-/Liouville theory is a dual description of the theory~(\ref{CShG-action}) with $M_0=0$. This gives the dimension of $\chi\bar\chi$, which results in~(\ref{M-M0-rel}).

Note that the spin (and hence, for generic values of parameters, the external spin) is a well\-/defined quantum number in any Lorentz invariant theory. In contrary, the scaling dimension is not a well\-/defined quantum number in a massive theory, since it is not an eigenvalue of a Hermitian operator. It is only an index that describes the decay of correlation functions at small distances. In particular, an admixture of operators of lower dimensions to a given operator does not change scaling dimension of the latter. Then the space $\cH^\kappa_{m\bm}$ is split into subspaces
$$
\cH^\kappa_{m\bm}=\bigoplus_{\Delta k\in\Z}\left(\cH^\kappa_{m\bm}\right)_{\Delta k},
$$
where $\left(\cH^\kappa_{m\bm}\right)_{\Delta k}$ is the subspace of all operators with external spin~$\Delta k$.
Each of these subspaces is organized as an infinite filtration by the total level~$L$:
$$
\left(\cH^\kappa_{m\bm}\right)_{\Delta k}\supset\cdots\supset\left(\cH^\kappa_{m\bm}\right)_{\Delta k,L}\supset\cdots
\supset\left(\cH^\kappa_{m\bm}\right)_{\Delta k,1}\supset\left(\cH^\kappa_{m\bm}\right)_{\Delta k,0}.
$$
Each element of the filtration $\left(\cH^\kappa_{m\bm}\right)_{\Delta k,L}$ consist of operators of the total level $\le L$. The last element of the filtration $\left(\cH^\kappa_{m\bm}\right)_{\Delta k,0}$ is one\-/dimensional, since it consist of the only exponential operator~$\phi^\kappa_{m\bm}$.

\subsection{\texorpdfstring{$S$}S matrix description}

In this paper we will develop the bootstrap form factor approach, whose starting point is an exact $S$ matrix. From the viewpoint of the scattering theory, the complex sinh\-/Gordon model is an integrable model with simple spectrum containing a particle and an antiparticle with mass~$M$ associated with the complex boson field~$\chi$. In two dimensions the standard parameterization of the momentum $P$ and energy $E$ is given in terms of the rapidity variable~$\theta$:
\eq$$
\Aligned{
P
&=M\sh\theta,
\\
E
&=M\ch\theta.
}\label{rapidity-def}
$$
Due to the integrability of the model, a multiparticle scattering matrix can be factorized into a product of two\-/particles ones. Let us denote the particle by a symbol $1$ and the antiparticle by~$\bar1$. The scattering matrix for two particles depends on the difference $\theta$ of their rapidities and is diagonal~\cite{Dorey:1994mg}:
\eq$$
S_{11}(\theta)=S_{\bar1\bar1}(\theta)
={\sh(\theta/2-\i\pi g)\over\sh(\theta/2+\i\pi g)},
\qquad
S_{1\bar1}(\theta)=S_{\bar11}(\theta)=S_{11}(i\pi -\theta).
\label{S-matrix}
$$
In the paper we will be interested in calculation of matrix elements of quasilocal operators with respect to eigenvectors of the Hamiltonian. These matrix elements are expressed in terms of meromorphic functions called form factors~\cite{Karowski:1978vz,Smirnov:1992vz}. These form factors can be found as solutions to a set of difference equations called form factor axioms. The data about the theory enters the form factor axioms through the $S$ matrix and, in this sense, the $S$ matrix completely describes the theory. Nevertheless, an important task is to identify the operators obtained in terms of solutions to the form factor axioms and those defined in the usual way, in terms of Lagrangian fields. Some results concerning this identification are presented in this paper.

Here we will follow the algebraic approach to form factors proposed in~\cite{Feigin:2008hs} and then developed in~\cite{Lashkevich:2013mca,Lashkevich:2013yja,Lashkevich:2014rua,Lashkevich:2014qna} for the sinh\-/Gordon model. A similar construction for the complex sinh\-/Gordon model was proposed in~\cite{Lashkevich:2016lzr}. Here we extend this construction by introducing the so called screening currents, which make it possible to obtain form factors of operators we are interested in.

\subsection{Relation to the \texorpdfstring{$\Z_N$}{ZN}~Ising models}

After a formal continuation of the parameter $g$ to the negative region $g=-N^{-1}$ for the values $N=2,3,\ldots$ the scattering matrix of the theory describes a basic particle and antiparticle scattering matrix of Koberle\--Swieca for the $\Z_N$~Ising models\cite{Koberle:1979sg}. There is an essential difference between the models however. In particular, for $N>3$ the $\Z_N$ Ising models not only contains the particles $1$ and $\bar1$, but also bound states of these fundamental particles. Form factors that involve bound state particles can be obtained from the form factors of the fundamental particles.

On the other hand, the $\Z_N$ Ising model can be treated as a perturbation of the Fateev\--Zamolodchikov $\Z_N$ conformal model with the Virasoro algebra central charge
\eq$$
c=2{N-1\over N+1}.
\label{ZN-centralcharge}
$$
This conformal model consists of a set of quasilocal primary operators $\phi^\kappa_{m\bm}(x)$ with $\kappa=0,1,\ldots,N$ and both $m,\bm$ equal to $\kappa$ modulo~$2$, and their $W_N$ descendants. These primary operators are in a one-to-one correspondence with the exponential operators defined in~(\ref{Phi-exp-def}). Let $\cS^0_{\Z_N}$ be the formal action of the conformal parafermion model. This action can be considered as a result of a reduction of the action~(\ref{CShG-action}) with $M_0=0$ and $g=-N^{-1}$. Then the formal action of the $\Z_N$ model reads
\eq$$
\cS_{\Z_N}=\cS^0_{\Z_N}-\lambda\int d^2x\,\phi^2_{00}(x).
\label{ZN-action}
$$
Due to the identification~(\ref{chi-barchi-V}) the second term can be considered as a counterpart of the mass term in the action~(\ref{CShG-action}), $\lambda\sim M_0^2$. The exact relation between the parameter $\lambda$ and the physical mass~$M$ was found in~\cite{Fateev:1993av}.

Note that the order parameter (`spin') operators $\sigma_k$, disorder parameter (`dual spin') operators $\mu_k$ and neutral `energy' density operators $\epsilon_k$ are identified with primary operators as follows:%
\footnote{With this notation $\chi\sim\sigma_{N-1}$, $\bar\chi\sim\sigma_1$, $\chi\bar\chi\sim\epsilon_1$.}
\eq$$
\sigma_k=\phi^k_{kk},
\qquad
\mu_k=\phi^k_{k,-k},
\qquad
\epsilon_k=\phi^{2k}_{00}.
\label{ZN-sigmamueps}
$$
We postpone discussing the reduction problem to another publication. Here we just mention, that the exponential operators are consistent with the reduction just for integer values of $\kappa,m,\bm$ subject to the conditions $m,\bm\equiv\kappa\pmod2$ and $|m|,|\bm|\le\kappa\le N$. These exponential operators represent primary operators of the $\Z_N$ Ising model. Besides, the conserved currents will be constructed here in such a way that they survive reduction and have a sense in the $\Z_N$ Ising model.

The identification with $\Z_N$ Ising models\cite{Fukuda:2001jd} essentially follows from the fact that the sine\-/Liouville model, i.e. the model described by the action (\ref{dual-action}) with $\lambda=0$, is nothing but the $SL(2,\R)/U(1)$ model of conformal field theory, a noncompact (or negative real $N$) counterpart of the $\Z_N$ symmetric conformal Fateev\--Zamolodchikov models, which are $SU(2)/U(1)$ coset conformal models.

In the case of the $\Z_2$ Ising model, form factors of primary and descendant operators were studied in~\cite{Cardy:1990pc}. The first results about primary operators in the $\Z_3$ Ising model were obtained in~\cite{Smirnov:1990pr}. Form factors of primary operators and a few descendant operators in general $\Z_N$ Ising models as well as their vacuum expectation values were studied within the algebraic approach in~\cite{Jimbo:2000ff,Fateev:2006js,Fateev:2009kp}. An alternative procedure based on a different approach was also proposed in~\cite{Babujian:2003za}. The construction proposed in~\cite{Lashkevich:2016lzr} can be continued to the $\Z_N$ Ising model, but for completeness demands a reduction procedure, which will be discussed elsewhere. Nevertheless, most of our results in this paper can be applied to the $\Z_N$ Ising model, and we will use the notation system adapted to it.

\subsection{Conserved currents}
\label{subsec-intr-intpert}

One of the advantages of our algebraic construction is that it makes it possible to find matrix elements of all operators including the descendant operators with both right and left chiralities. This question turns out to be rather nontrivial in other approaches starting from the pioneer work\cite{Koubek:1993ke}. An important understanding concerning the equivalence of the operator contents of conformal models and their massive perturbations was achieved in~\cite{Jimbo:2003ge,Jimbo:2003ne,Delfino:2007bt}.

In this paper we pay a special attention to form factors of physically interesting spin $k\in\Z$ descendants $T_k$ and $\Theta_k$ of the unit operator and of the operator $\phi^2_{00}$ respectively. These operators are components of conserved currents and define the local integrals of motion:
\eq$$
\Aligned{
I_k
&=\int{dz\over2\pi}\,T_{k+1}(x)+\int{d\bz\over2\pi}\,\Theta_{k-1}(x),
\\
I_{-k}
&=\int{d\bz\over2\pi}\,T_{-k-1}(x)+\int{dz\over2\pi}\,\Theta_{-k+1}(x),
\qquad k=1,2,\ldots.
}\label{Ik-in-currents}
$$
The subscript $k$ of $I_k$ is usually called the spin of the current in the following sense: if $\cO(x)$ is a spin $s$ operator, the operator $[I_k,\cO(x)]$ is a spin $s+k$ operator.

The existence of commuting local integrals of motion provides the integrability of the massive model\cite{Zamolodchikov:1989zs}. In the off\-/critical theory the following continuity equations hold for these currents
\eq$$
\Aligned{
\d_-T_k
&=\d_+\Theta_{k-2},
\\
\d_+T_{-k}
&=\d_-\Theta_{-k+2}.
}\label{continuity}
$$
In this paper we obtain multiparticle form factors of the currents as well as those of the combined operators~$T_kT_{-l}$ ($k,l\ge2$) defined according to Zamolodchikov's rule\cite{Zamolodchikov:2004ce,Cardy:2018jho}:
\eq$$
T_kT_{-l}(x)
=\lim_{\epsilon\to 0}\left(T_k(x+\epsilon)T_{-l}(x)-\Theta_{k-2}(x+\epsilon)\Theta_{2-l}(x)
-\d_\mu J^\mu_{kl}(x,\epsilon)\right),
\label{TT-def}
$$
where the last term is a combination of space\-/time derivatives that cancels the singular part of operators products in the first two terms. Thus this operator is defined modulo space\-/time derivatives.

Form factors of the operators $T_kT_{-l}(x)$ ($k,l\ge2$) where first discussed (for $k=l=2$) by G.~Delfino and G.~Niccoli in~\cite{Delfino:2004vc,Delfino:2005wi,Delfino:2006te}. Recently~\cite{Lashkevich:2014qna} we obtained form factors of the products with arbitrary $k,l$ in the sinh\-/Gordon model and the perturbed minimal $M(2,2n+1)$ model. Here we show how to transfer this method to the complex sinh\-/Gordon model and propose explicit expressions for form factors
$$
\langle\vac|\cO|\theta_1,\ldots,\theta_n\rangle_{\alpha_1\ldots\alpha_N},
\qquad
\cO=T_k,\>\Theta_{k-2},\>T_{-l},\,\Theta_{2-k},\>T_kT_{-l}.
$$
This result was achieved by studying algebraic structures that arise in the free field realization of form factors.

\subsection{Integrable current--current perturbations}

The operators $T_kT_{-k}$ are interesting since they produce a family of irrelevant integrable deformations of integrable quantum field theories\cite{Zamolodchikov:1991vx,Mussardo:1999aj,Zamolodchikov:2006pc}. Recently, F.~Smirnov and A.~Zamolodchikov proved\cite{Smirnov:2016lqw} the integrability of infinite\-/dimensional families of such theories with the action of the form
\eq$$
\cS_\lambda
=\cS_0-\sum^\infty_{k=2}\lambda_{2k-1}\int d^2x\,[T_{2k}T_{-2k}](x),
\label{SZ-action}
$$
where the formal action $\cS_0$ describes any two\-/dimensional integrable model of QFT. The bracket symbol is defined as
\eq$$
[T_kT_{-l}](x)
=T_kT_{-l}(x)+\delta_{k,2}\langle\Theta_0(0)\rangle\,\Theta_{2-l}(x)
+\delta_{l,2}\langle\Theta_0(0)\rangle\,\Theta_{k-2}(x)
-\delta_{k,2}\delta_{l,2}\langle\Theta_0(0)\rangle^2.
\label{TTred-def}
$$
This modification for $k=2$ or $l=2$ is necessary to avoid divergences in the diagonal matrix elements, which enter the perturbation theory for the $S$ matrix. It was assumed that the structure of the spectrum and the $S$ matrix of the initial model forbids integrals of motion of even spin. It was conjectured that for a massive integrable model the resulting $S$ matrix takes a general form%
\footnote{Though the operator $T_kT_{-k}$ is not uniquely defined, the corresponding integral over the plane is well\-/defined since the space\-/time derivatives produce zero contribution.}
\eq$$
S_\lambda(\theta)=S(\theta)\e^{\i\Phi_\lambda(\theta)},
\qquad
\Phi_\lambda(\theta)=-\sum^\infty_{k=1}\lambda_{2k-1}\left(Z_{2k}\over M\right)^2\sh(2k-1)\theta,
\label{S-SZ-gen}
$$
where $S(\theta)$ is the $S$ matrix of the underlying local integrable model and
\eq$$
Z_k=\e^{-k\theta}\langle\theta|T_k(0)|\theta\rangle
\label{Zk-def}
$$
are constants. The conjecture was checked in~\cite{Smirnov:2016lqw} for the soliton $S$ matrix of the sine\-/Gordon theory by means of the fermion representation~\cite{Jimbo:2011bc} in the first order, and it was conjectured that the result is exact. Independently, we checked the conjecture for the sinh\-/Gordon model and, therefore, for the first breather $S$ matrix of the sine\-/Gordon model in~\cite{Lashkevich:2016gev} by means of the algebraic construction we developed earlier.

\subsection{Perturbations of the complex sinh-Gordon model and the \texorpdfstring{$\Z_N$}{ℤN} Ising model}

In the complex sinh\-/Gordon model the scattering does not exclude even spin integrals of motion, so that there is just one integral of motion for every nonzero spin. In the present paper we obtain the CDD factor in this case. Namely, we consider the model
\eq$$
\cS_\lambda[\chi,\bar\chi]
=\int{d^2x\over4\pi}\,\left({\d_\mu\chi\,\d^\mu\bar\chi\over1+g_0\chi\bar\chi}
-M_0^2\chi\bar\chi\right)-\sum^\infty_{k=1}\lambda_k\int d^2x\,[T_{k+1}T_{-k-1}](x).
\label{CShG-pert}
$$

An important part of the derivation of the exact CDD factor~\cite{Smirnov:2016lqw} was a conjecture
on the structure of the $1+1\to1+1$ matrix elements of the operators $T_kT_{-k}$. In our earlier work \cite{Lashkevich:2016gev} we were able to provide an analytic computation of these matrix elements in the case of the sinh\-/Gordon model. In the present work we develop a similar technique to derive the expression for the complex sinh\-/Gordon model. The result is
\eq$$
\Aligned{
\lim_{\theta_3\to\theta_1\atop\theta_4\to\theta_2}
{}_{11}\langle\theta_1\theta_2|[T_kT_{-k}](0)|\theta_3\theta_4\rangle_{11}
&=4Z_k^2\sh(\theta_1-\theta_2)\sh(k-1)(\theta_1-\theta_2),
\\
\lim_{\theta_3\to\theta_1\atop\theta_4\to\theta_2}
{}_{1\bar1}\langle\theta_1\theta_2|[T_kT_{-k}](0)|\theta_3\theta_4\rangle_{1\bar1}
&=(-)^k4Z_k^2\sh(\theta_1-\theta_2)\sh(k-1)(\theta_1-\theta_2),
}\label{TT-ff}
$$
This formula generalizes the formula by Smirnov and Zamolodchikov to the case of even currents. Following the general argument of~\cite{Smirnov:2016lqw} we propose that the exact scattering matrix is given by the expression 
\eq$$
S_{11}^\lambda(\theta)=S_{\bar1\bar1}^\lambda(\theta)
=S_{11}(\theta)\e^{\i\Phi^\lambda_+(\theta)},
\qquad
S_{1\bar1}^\lambda(\theta)=S_{\bar11}^\lambda(\theta)
=S_{1\bar1}(\theta)\e^{\i\Phi^\lambda_-(\theta)},
\label{S-lambda-gen}
$$
where
\eq$$
\Phi^\lambda_\ve(\theta)=-\sum^\infty_{k=1}\ve^{k+1}\lambda_k\left(Z_{k+1}\over M\right)^2\sh k\theta.
\label{Phi-gen}
$$
We find a remarkable agreement with the Smirnov\--Zamolodchikov proposal for the modified scattering matrix. To calculate the matrix elements~(\ref{TT-ff}) we provide a construction for the currents $T_k,\Theta_k$ and, which demands a heuristic step, for the products~$T_kT_{-l}$.

As an immediate consequence, we can also propose that the corresponding $\Z_N$ Ising models also have a similar integrable perturbation 
\eq$$
\cS_\lambda=\cS_{\Z_N}-\lambda \int d^2x\,\phi_{00}^2-\sum^\infty_{k=1\atop k\not\equiv0\bmod N}
\lambda_k\int d^2x\,[T_{k+1}T_{-k-1}].
\label{ZN-pert}
$$
The CDD factors arising in this case can be extracted expression~(\ref{Phi-gen}) for $g=-N^{-1}$. In the $\Z_N$ Ising model there are $N-1$ particles of masses
\eq$$
M_s=M{\sin{\pi s\over N}\over\sin{\pi\over N}},
\qquad
s=1,\ldots,N-1.
\label{Mk-def}
$$
Let $S_{ss'}(\theta)$ be the $S$ matrix of particles $s$ and $s'$ in the $\Z_N$ Ising model. Then for the $S$ matrix of the model~(\ref{ZN-pert}) we have
\eq$$
S^\lambda_{ss'}(\theta)
=S_{ss'}(\theta)\e^{\i\Phi^\lambda_{ss'}(\theta)},
\label{S-lambda-ZN}
$$
where
\eq$$
\Phi^\lambda_{ss'}(\theta)
=-\sum^\infty_{k=1}\lambda_k\left(Z_{k+1}\over M\right)^2\fr_{sk}\fr_{s'k}\sh k\theta,
\qquad
\fr_{sk}={\sin{\pi sk\over N}\over\sin{\pi k\over N}}.
\label{Phi-ZN}
$$
This expression is obtained from (\ref{S-lambda-gen}), (\ref{Phi-gen}) by fusion of particles starting from $\Phi^\lambda_{11}(\theta)=\Phi^\lambda_+(\theta)$. As a consistency check we may notice that $\Phi^\lambda_{N-1,1}(\theta)=\Phi^\lambda_-(\theta)$, since the particle $(N-1)$ is the antiparticle to the particle~$1$.

\subsection{Lorentz non-invariant perturbations}

The action (\ref{CShG-pert}) admits a generalization
\eq$$
\cS_\lambda[\chi,\bar\chi]
=\int{d^2x\over4\pi}\,\left({\d_\mu\chi\,\d^\mu\bar\chi\over1+g_0\chi\bar\chi}
-M_0^2\chi\bar\chi\right)-\sum^\infty_{k,l=1}\lambda_{kl}\int d^2x\,[T_{k+1}T_{-l-1}](x).
\label{CShG-LNIpert}
$$
The operators $[T_kT_{-l}]$ has spin $k-l$ so that the contributions with $k\ne l$ are Lorentz non\-/invariant. Nevertheless, the corresponding $1+1\to1+1$ matrix elements also have a simple form:
\eq$$
\Aligned{
\lim_{\theta_3\to\theta_1\atop\theta_4\to\theta_2}
{}_{11}\langle\theta_1\theta_2|[T_kT_{-l}](0)|\theta_3\theta_4\rangle_{11}
&=4Z_kZ_l\sh(\theta_1-\theta_2)\fs^+_{k-1,l-1}(\theta_1,\theta_2),
\\
\lim_{\theta_3\to\theta_1\atop\theta_4\to\theta_2}
{}_{1\bar1}\langle\theta_1\theta_2|[T_kT_{-l}](0)|\theta_3\theta_4\rangle_{1\bar1}
&=4Z_kZ_l\sh(\theta_1-\theta_2)\fs^-_{k-1,l-1}(\theta_1,\theta_2)
}\label{TT-LNIff}
$$
with
\eq$$
\fs^\ve_{kl}(\theta_1,\theta_2)
={1\over2}\left(\ve^{k+1}\e^{k\theta_1-l\theta_2}-\ve^{l+1}\e^{k\theta_2-l\theta_1}\right).
\label{fraks-def}
$$
Relying on this result, we may conjecture that the $S$ matrices, which now depend on the two rapidities, read
\eq$$
\Aligned{
S_{11}^\lambda(\theta_1,\theta_2)=S_{\bar1\bar1}^\lambda(\theta_1,\theta_2)
&=S_{11}(\theta_1-\theta_2)\e^{\i\Phi^\lambda_+(\theta_1,\theta_2)},
\\
S_{1\bar1}^\lambda(\theta_1,\theta_2)=S_{\bar11}^\lambda(\theta_1,\theta_2)
&=S_{1\bar1}(\theta_1-\theta_2)\e^{\i\Phi^\lambda_-(\theta_1,\theta_2)},
}\label{S-lambda-LNIgen}
$$
where the phase shift, which generalizes~(\ref{Phi-gen}), is
\eq$$
\Phi^\lambda_\ve(\theta_1,\theta_2)
=-\sum^\infty_{k,l=1}\lambda_{kl}{Z_{k+1}Z_{l+1}\over M^2}\fs^\ve_{kl}(\theta_1,\theta_2).
\label{Phi-LNIgen}
$$

For the Lorentz non\-/invariant action for the $\Z_N$ Ising model
\eq$$
\cS_\lambda=\cS_{\Z_N}-\lambda\int d^2x\,\phi_{00}^2-\sum^\infty_{k=1\atop k\not\equiv0\bmod N}
\lambda_{kl}\int d^2x\,[T_{k+1}T_{-l-1}]
\label{ZN-LNIpert}
$$
we obtain the following:
\eq$$
S^\lambda_{ss'}(\theta_1,\theta_2)
=S_{ss'}(\theta)\e^{\i\Phi^\lambda_{ss'}(\theta_1,\theta_2)},
\label{S-lambda-ZN-LNI}
$$
where
\eq$$
\Phi^\lambda_{ss'}(\theta_1,\theta_2)
=-\sum^\infty_{k,l=1}\lambda_{kl}{Z_{k+1}Z_{l+1}\over2M^2}
\left(\fr_{sk}\fr_{s'l}\e^{k\theta_1-l\theta_2}-\fr_{sl}\fr_{s'k}\e^{k\theta_2-l\theta_1}\right).
\label{Phi-ZN-LNI}
$$

\subsection{Structure of the paper}

The paper is organized as follows. In Sect.~\ref{sec-freefield} we recall the basic facts about form factors and describe the free field construction introduced in~\cite{Lashkevich:2016lzr}. In Sect.~\ref{sec-screening} we describe the new objects, the so called screening currents, analogous to those introduced in~\cite{Lashkevich:2013yja,Lashkevich:2014qna} for the sinh\-/Gordon model. In fact, this realization makes it possible to calculate form factors in a rather direct way. In Sect.~\ref{sec-conscur} we obtain free field realizations for the conserved currents $T_k,\Theta_k$. In Sect.~\ref{sec-TT} we consider the products~$T_kT_{-l}$. Obtaining a compact form for the two-by-two diagonal matrix elements of $T_kT_{-l}$ for arbitrary $k,l$ demands some tricks, which are described there.

\section{Free field construction for form factors}
\label{sec-freefield}

\subsection{Form factor axioms}
\label{sec-ff-axioms}
Our first aim is to formulate briefly the free field construction for form factors of physical operators in the complex sinh\-/Gordon model. A natural bases in the spaces of operators consist of exponential operators~$\phi^\kappa_{m\bm}(x)$ and their descendants of the form~(\ref{descendant-def}). In the form factor approach all quasilocal operators $\cO$ are characterized by an infinite set of matrix elements in the basis of asymptotic states. Due to the crossing symmetry it is sufficient to describe the matrix elements between the vacuum and incoming particles only. Let $|\theta_1,\ldots,\theta_n\rangle_{\alpha_1\ldots\alpha_n}$ be the in\-/state of $n$ particles with the velocities $\theta_i$ and internal states $\alpha_i=1,\bar1$. Consider any local or quasilocal operator~$\cO(x)$. The meromorphic functions $F_\cO(\theta_1,\ldots,\theta_n)$ of complex variables $\theta_i$, whose values on the real axis are given by
\eq$$
F_\cO(\theta_1,\ldots,\theta_n)_{\alpha_1\ldots\alpha_n}
=\langle\vac|\cO(0)|\theta_1\ldots\theta_n\rangle_{\alpha_1\ldots\alpha_n},
\quad\text{if }
\theta_1>\ldots>\theta_n,
\label{FF-def}
$$
are called form factors of the operator~$\cO$. Every correlation function can be expressed in terms of form factors due to the spectral expansion (see~\cite{Smirnov:1992vz} and references therein for details). The form factors must be neutral with respect to the charge $Q$ defined in~(\ref{Qcharge-Phi}):
\eq$$
Q(\cO)+\sum^n_{i=1}\alpha_i=0.
\label{ff-neutrality}
$$

Form factors are shown to satisfy a set of difference equations called form factor axioms~\cite{Smirnov:1992vz}. Let $s_\cO$ be the Lorentz spin of the operator $\cO(x)$ and $\gamma_\cO(\alpha)$ be a mutual locality index of the operator $\cO$ and a bosonic `elementary field' operator that corresponds to the annihilation of a particle of type~$\alpha$, in our case $\chi$ for $\alpha=1$ and $\bar\chi$ for $\alpha=\bar1$. In the case of a diagonal $S$ matrix, which is the case of our interest here, the form factor axioms read
\subeq{\label{ffax}
\Align$$
F_\cO(\theta_1+\vartheta,\ldots,\theta_n+\vartheta)_{\alpha_1\ldots\alpha_n}
&=\e^{s_\cO\vartheta}F_\cO(\theta_1,\ldots,\theta_n)_{\alpha_1\ldots\alpha_n},
\label{ffax-spin}
\\
F_\cO(\theta_1,\theta_2,\ldots,\theta_n)_{\alpha_1\alpha_2\ldots\alpha_n}
&=\e^{2\pi\i\gamma_\cO(\alpha_1)}F(\theta_2,\ldots,\theta_n,\theta_1-2\pi\i)_{\alpha_2\ldots\alpha_n\alpha_1},
\label{ffax-cyclic}
\\
F_\cO(\ldots,\theta_i,\theta_{i+1},\dots)_{\ldots\alpha_i\alpha_{i+1}\ldots}
&=S_{\alpha_i\alpha_{i+1}}(\theta_i-\theta_{i+1})
F_\cO(\ldots,\theta_{i+1},\theta_i,\ldots)_{\ldots\alpha_{i+1}\alpha_i\ldots},
\label{ffax-commut}
\\
\Res_{\vartheta'=\vartheta+\i\pi}
F_\cO(\vartheta',\vartheta,\theta_1,\ldots,\theta_n)_{\bar\alpha\alpha\alpha_1\ldots\alpha_n}
\notag
\\
&\hskip -6em
=-\i\left(1-\e^{2\pi\i\gamma_\cO(\alpha)}\prod^n_{i=1}S_{\alpha\alpha_i}(\vartheta-\theta_i)\right)
F_\cO(\theta_1,\ldots,\theta_n)_{\alpha_1\ldots\alpha_n}.
\label{ffax-kinematic}
$$}
The last equation fixes the residue of the kinematic pole of form factors on the physical sheet. In the presence of bound states one needs to add equations for the residues of dynamic poles.

For a diagonal $S$ matrix the form factor can be factorized into the product of a special solution $R_{n,n'}$ to equations (\ref{ffax-cyclic}),~(\ref{ffax-commut}) and a function $J_\cO$ with simple analytic properties. Namely, we have
\Align$$
F_\cO(\theta_1,\ldots,\theta_n|\theta'_1,\ldots,\theta'_{n'})
&\equiv F_\cO(\theta_1,\ldots,\theta_n,\theta'_1,\ldots,\theta'_{n'})
_{\underbrace{1\ldots1}_n\underbrace{\bar1\ldots\bar1}_{n'}}
\notag
\\
&=J_\cO(\e^{\theta_1},\ldots,\e^{\theta_n}|\e^{\theta'_1},\ldots,\e^{\theta'_{n'}})
\notag
\\*
&\quad\times
R_{n,n'}(\theta_1,\ldots,\theta_n|\theta'_1,\ldots,\theta'_{n'}).
\label{ff-JRfactor}
$$
The function $J_\cO(X|Y)\equiv J_\cO(x_1,\ldots,x_n|y_1,\ldots,y_{n'})$ has the form
\eq$$
J_\cO(X|Y)=\prod^n_{i=1}x_i^{\gamma_\cO(1)}\prod^{n'}_{j=1}y_j^{\gamma_\cO(\bar1)}
\check J_\cO(X|Y),
\label{JcO-form}
$$
where $\check J_\cO(X|Y)$ is a rational function symmetric in the variables $x_i$ and in the variables $y_i$ separately. The set of functions $J_\cO(X|Y)$ contain the complete information about the physical operator $\cO$ and are the main object to be studied in this work. In what follows we describe a free field construction for these function, which solves the form factor axioms for $F_\cO$ and, as we expect, makes it possible, in principle, to obtain form factors of any quasilocal operator of the form~(\ref{descendant-def}).

The second factor $R_{n,n'}$ is already fixed in terms of the known functions and therefore will not be so interesting. Its explicit form is~\cite{Oota:1995ga}
\eq$$
R_{n,n'}(\theta_1,\ldots,\theta_n|\theta'_1,\ldots,\theta'_{n'})
=C_N^{n+n'}\prod^n_{i<j}R(\theta_i-\theta_j)\prod^{n'}_{i<j}R(\theta'_i-\theta'_j)
\prod^n_{i=1}\prod^{n'}_{j=1}\Rdag(\theta_i-\theta'_j),
\label{fkmbmdef}
$$
where
\eq$$
C_N={1\over\sqrt{2\sin\pi g}}={\i\over\sqrt{2\sin{\pi\over N}}}.
\label{CN-def}
$$
The functions $R,\Rdag$ read
\eq$$
\Gathered{
R(\theta)={\sh{\theta\over2}\over\sh\left({\theta\over2}-{\i\pi\over N}\right)}\bar R(\theta),
\qquad
\Rdag(\theta)=\bar R^{-1}(\theta-\i\pi),
\\
\bar R(\theta)
=\prod^\infty_{n=1}{\Gamma\left({\i\theta\over2\pi}-{1\over N}+n\right)
  \Gamma\left(-{\i\theta\over2\pi}-{1\over N}+n\right)\Gamma^2\left(n+{1\over N}\right)
  \over\Gamma\left({\i\theta\over2\pi}+{1\over N}+n\right)
  \Gamma\left(-{\i\theta\over2\pi}+{1\over N}+n\right)\Gamma^2\left(n-{1\over N}\right)},
}\label{RRdag-def}
$$
and satisfy the equations
$$
\Aligned{
R(2\pi\i-\theta)
&=R(\theta),
\quad
&R(\theta)
&=S_{11}(\theta)R(-\theta),
\\
\Rdag(2\pi\i-\theta)
&=\Rdag(\theta),
\quad
&\Rdag(\theta)
&=S_{1\bar1}(\theta)\Rdag(-\theta).
}
$$
Here and below we always use the parameter $N=-g^{-1}$, since it is more convenient in our technique. We will not assume it to be an integer. Moreover, we will usually assume it irrational and negative, as it is necessary in the complex sinh\-/Gordon model. Nevertheless, the results are applicable to the $\Z_N$ Ising model for those operators that are compatible with the reduction.

It is straightforward to check that the form factors are consistently defined by axiom~(\ref{ffax-commut}), so that they satisfy axioms~(\ref{ffax}a,b) for any set of homogeneous rational functions $\check J_\cO(X|Y)$, while the kinematic pole condition (\ref{ffax-kinematic}) reads
\eq$$
J_\cO(X,z'|Y,z)
=-\omega^{\pm{m-\bm\over4}}{(\omega^{1/2}-\omega^{-1/2})z\over z'+z}
D_{m\bm}(X|Y|z)J_\cO(X|Y)+O(1),
\qquad
z'\to\e^{\pm\i\pi}z.
\label{JcO-kinpole}
$$
Here
\eq$$
D_{m\bm}(X|Y|z)=\omega^{m-\bm\over4}\prod_{x\in X}f_+\left(-{z\over x}\right)\prod_{y\in Y}f_-\left(z\over y\right)
-\omega^{\bm-m\over4}\prod_{x\in X}f_-\left(-{z\over x}\right)\prod_{y\in Y}f_+\left(z\over y\right)
\label{Drs-def}
$$
with
\eq$$
f_\ve(z)={\omega^{\ve/2}z-\omega^{-\ve/2}\over z-1},
\qquad
\omega=\e^{2\pi\i/N}.
\label{f-def}
$$
The functions $f_\ve(z)$ will play an important role in the free field construction.

\subsection{Oscillators and currents}

To describe from factors of exponential and descendant operators in the complex sinh\-/Gordon model we follow the free field approach proposed in~\cite{Feigin:2008hs}. In this method the functions $J_\cO$ from~(\ref{ff-JRfactor}) are represented as matrix elements of free fields of a special form. In this section we recall the results of~\cite{Lashkevich:2016lzr} with a slight modification necessary to introduce screening currents in the next section.

First, introduce some notation. In what follows we will use the parameter $\omega$ defined in (\ref{f-def}). Sometimes it will be convenient to use the `conjugate' parameter
\eq$$
\tomega=\e^{-\i\pi}\omega^{-1}=\e^{-\i\pi(N+1)/N}.
\label{tomega-def}
$$
Besides, for brevity we will use the notation
\eq$$
\Aligned{{}
\brks{k}
&=\omega^{k/2}-\omega^{-k/2}=2\i\sin{\pi k\over N},
\quad
&\tbrks{k}
&=\tomega^{k/2}-\tomega^{-k/2}=-2\i\sin{\pi(N+1)k\over N},
\\
\brcs{k}
&=\omega^{k/2}+\omega^{-k/2}=2\cos{\pi k\over N},
\quad
&\tbrcs{k}
&=\omega^{k/2}+\omega^{-k/2}=2\cos{\pi(N+1)k\over N}.
}\label{brackets-def}
$$

The corresponding bosonization procedure is formulated in terms of two families of oscillators $\alpha^\pm_k$ and $\beta^\pm_k$ with $k\in\Z$, $k\ne0$ and the oscillator $\gamma^\pm$. We assume the nonzero commutation relations for these elements to be
\eq$$
\Aligned 
{
&[\alpha^{\ve}_k,\alpha^{\ve'}_l]=kA^{(1)}_{{\ve}k}
\delta_{{\ve}+{\ve'},0}\delta_{k+l,0},
\\
&[\beta^{\ve}_k,\beta^{\ve'}_l]=kA^{(2)}_{{\ve}k}\delta_{{\ve}+{\ve'},0}\delta_{k+l,0},
\\
&[\gamma^-,\gamma^+]=(\log\omega)^{-1}.}
\label{ddcommut}
$$
The numerical coefficients $A^{(i)}_k$ ($i\in\{1,2\}$), which enter the definition of commutation relation of the oscillators, are taken in the form
\eq$$
\Aligned{
A^{(i)}_k
={1\over2}(\omega^{k/2}-\omega^{-k/2})(\omega^{k/2}+(-1)^{k+i}\omega^{-k/2}).
}\label{ABdef}
$$
We also need the zero mode operators $P,\Pdag$ and the operators $\ud,\uddag$, which are conjugated to them:
\eq$$
[\ud,P]=[\uddag,\Pdag]=1,
\qquad
[\ud,\Pdag]=[\uddag,P]=0.
\label{dP-commut}
$$
The zero mode operators commute with all the oscillators $\alpha^\pm_k,\beta^\pm_k,\gamma^\pm$.

Define the vacuum vectors $|1\rangle_{rs}$ for these auxiliary oscillators:
\eq$$
\Aligned{
\alpha^\ve_k|1\rangle_{rs}
&=\beta^\ve_k|1\rangle_{rs}=0\quad(k>0,\ve=\pm),
\\
P|1\rangle_{rs}
&=p|1\rangle_{rs}=(2r+N-1)|1\rangle_{rs},
\\
\Pdag|1\rangle_{rs}
&=\pdag|1\rangle_{rs}=(2s+N-1)|1\rangle_{rs}.
}\label{1vacsdef}
$$
Here $r,s$ are two real numbers that fix the eigenvalues of the operators~$P,\Pdag$. The operators $\alpha^\pm_{-k}$, $\beta^\pm_{-k}$ ($k>0$) generate the Fock space, which will be denoted by~$\bcF_{rs}$. The conjugate vacuum vectors ${}_{rs}\langle1|$ are defined similarly:
\eq$$
\Aligned{
{}_{rs}\langle1|\alpha^{\ve}_{-k}
&={}_{rs}\langle1|\beta^{\ve}_{-k}=0 \quad(k>0,\ve=\pm),
\\
{}_{rs}\langle1|P
&={}_{rs}\langle1|p={}_{rs}\langle1|(2r+N-1),
\\
{}_{rs}\langle1|\Pdag
&={}_{rs}\langle1|\pdag={}_{rs}\langle1|(2s+N-1).
}\label{1vacsdefconj}
$$
We will assume ${}_{rs}\langle1|1\rangle_{rs}=1$. The operators $\alpha^\pm_k$, $\beta^\pm_k$ ($k>0$) generate another Fock space indicated by~$\cF_{rs}$. The exponents of the operators $\ud,\uddag$ naturally shift the parameters $r,s$:
\eq$$
\e^{2a\ud+2b\uddag}|1\rangle_{rs}=|1\rangle_{r-a,s-b}.
\label{dddag-exp}
$$

As for the operators $\gamma^{\ve}$, we will demand
\eq$$
{}_{rs}\langle1|\omega^{a\gamma^-+(b-a)\gamma^+}|1\rangle_{rs}=1.
\label{gamma-vacmatel}
$$
In fact, it means that the vacuum vectors depend on the value of~$b$. Nevertheless, since for any operator $\cO$ this value turns out to be fixed, $b=-Q(\cO)$, we may omit this parameter.

Introduce the normal ordering in consistency with the above definitions:
\eq$$
\lcolon\alpha^{\ve}_k\alpha^{\ve'}_{-k}\rcolon
=\alpha^{\ve'}_{-k}\alpha^{\ve}_k,
\qquad
\lcolon\beta^\ve_k\beta^{\ve'}_{-k}\rcolon
=\beta^{\ve'}_{-k}\beta^\ve_k,
\qquad
k>0.
\label{alphabeta-normal}
$$
The normal ordering prescription for zero mode operators stands that we put $P,\Pdag$ to the right of $\ud,\uddag$ and collect exponents of $\gamma^\ve$:
\eq$$
\lcolon\prod_i\omega^{\mu_i\gamma^-+\nu_i\gamma^+}\rcolon=\omega^{\gamma^-\sum_i\mu_i+\gamma^+\sum_i\nu_i}.
\label{gamma-normal}
$$
Define the vertex operators
\eq$$
\Aligned {
\lambda_\ve(z)
&=\omega^{\gamma^\ve}\exp\sum_{k\ne0}{\left(\alpha^\ve_k+\beta^\ve_k\right)}{z^{-k}\over k},
\\
\lambdadag_\ve(z)
&=\omega^{-\gamma^\ve}\exp\sum_{k\ne0}{\left(\alpha^\ve_k-\beta^\ve_k\right)}{z^{-k}\over k}.
\label{lambdadef}
}
$$
The normal order sign is unnecessary here, because all operators in the exponents commute with each other. Since $\lambda_\ve(x),\lambdadag_\ve(x)$ do not contain the zero modes, their vacuum correlation functions are $r$ and $s$ independent and will be simply designated as $\langle\cdots\rangle$. Due to the Wick theorem we have
\eq$$
\phi_1\cdots\phi_n=\langle\phi_1\cdots\phi_n\rangle\lcolon\phi_1\cdots\phi_n\rcolon,
\label{expprod-norm}
$$
where $\phi_i$ are arbitrary normally ordered exponents of the oscillators $\alpha^\ve_k,\beta^\ve_k,\gamma^\ve$ like $\lambda_\ve(z)$ or $\lambdadag_\ve(z)$. The vacuum average $\langle\phi_1\cdots\phi_n\rangle$ factorizes into pair correlation functions:
\eq$$
\langle\phi_1\cdots\phi_n\rangle
=\prod^n_{i<j}\langle\phi_i\phi_j\rangle.
\label{expprod-av}
$$
The pair correlation functions for $\lambda_\ve(z),\lambdadag_\ve(z)$ are given by
\eq$$
\Gathered{
\langle\lambda_\ve(z')\lambda_\ve(z)\rangle=\langle\lambdadag_\ve(z')\lambdadag_\ve(z)\rangle
=\langle\lambda_\ve(z')\lambdadag_\ve(z)\rangle=\langle\lambdadag_\ve(z')\lambda_\ve(z)\rangle=1,
\\
\Aligned{
\langle\lambda_\ve(z')\lambda_{-\ve}(z)\rangle=\langle\lambdadag_\ve(z')\lambdadag_{-\ve}(z)\rangle
&=f_\ve\left(z\over z'\right),
\\
\langle\lambda_\ve(z')\lambdadag_{-\ve}(z)\rangle=\langle\lambdadag_\ve(z')\lambda_{-\ve}(z)\rangle
&=f_{-\ve}\left(-{z\over z'}\right),
}}\label{lambdapair-av}
$$
where $f_\ve(z)$ is defined in~(\ref{f-def}). Notice that all the operators $\lambda$ commute with each other:
\eq$$
\Gathered{
\lambda_{\ve'}(z')\lambda_\ve(z)=\lambda_\ve(z)\lambda_{\ve'}(z'),
\qquad
\lambdadag_{\ve'}(z')\lambdadag_\ve(z)=\lambdadag_\ve(z)\lambdadag_{\ve'}(z'),
\qquad
\text{if $z'\ne z$;}
\\
\lambda_{\ve'}(z')\lambdadag_\ve(z)=\lambdadag_\ve(z)\lambda_{\ve'}(z'),
\qquad
\text{if $z'\ne-z$.}
}\label{lambda-commut}
$$
The exception points $z'=\pm z$ are related to the poles of the pair correlation functions~(\ref{lambdapair-av}). The pole contributions from $\langle\lambda_\pm\lambda_\mp\rangle$ and $\langle\lambdadag_\pm\lambdadag_\mp\rangle$, which break the commutativity in the first line, will cancel each other in final expressions. The contribution from $\langle\lambda_\pm\lambdadag_\mp\rangle$, which breaks the commutativity in the second line, is important and is responsible for the kinematic poles of form factors. It reads:
\eq$$
\lambda_\ve(z')\lambdadag_{-\ve}(z)
=-\ve{\brks{1}z\over z'+z}s_\ve(z)+O(1)
\quad\text{as }z'\to-z.
\label{lambdalambdadagpole}
$$
The operators $s_\ve(z)$ are defined as
\eq$$
s_\ve(z)=\lcolon\lambda_\ve(-z)\lambdadag_{-\ve}(z)\rcolon
=\omega^{\ve(\gamma^+-\gamma^-)}\lcolon\exp\left(\sum_{k\ne0}{D^\ve_k\over k}z^{-k}\right)\rcolon,
\label{spmdef}
$$
where
\eq$$
D^\ve_k=(-)^k(\alpha^\ve_k+\beta^\ve_k)+\alpha^{-\ve}_k-\beta^{-\ve}_k,
\label{Dpmdef}
$$
Important properties of $s_\ve(z)$ are
\eq$$
\Aligned{
\langle s_{\ve'}(z')\lambda_\ve(z)\rangle
&=f_{\ve'}\left(-{z\over z'}\right),
\\
\langle s_{\ve'}(z')\lambdadag_\ve(z)\rangle
&=f_{-\ve'}\left(z\over z'\right),
}\label{slambda-pair}
$$
and
\eq$$
\langle s_{\ve'}(z')s_\ve(z)\rangle
=f_+\left(z\over z'\right)f_-\left(z\over z'\right).
\label{ss-pair}
$$

Now we are ready to define the vertex operators $t(z)$, $\tdag(z)$, which correspond in the free field construction to the operators that create the particles:%
\footnote{The definition here differs from that of~\cite{Lashkevich:2016lzr} by non\-/rational prefactors, which will drastically simplify the commutation relations with the screening currents and screening operators later.}
\eq$$
\Aligned{
t(z)
&=z^{(P-\Pdag)/2N}\left(\omega^{P/4}\lambda_+(z)+\omega^{-P/4}\lambda_-(z)\right),
\\
\tdag(z)
&=z^{(\Pdag-P)/2N}\left(\omega^{\Pdag/4}\lambdadag_+(z)+\omega^{-\Pdag/4}\lambdadag_-(z)\right).
}\label{ttdag-def}
$$
These operators are not pure exponents, and we must decompose the corresponding products before applying the rules (\ref{expprod-norm}),~(\ref{expprod-av}). Moreover, they contain zero modes, so that it is important to specify the modules $\bcF_{rs}$, in which they act.

From the property (\ref{lambdalambdadagpole}) we easily obtain
\eq$$
t(z')\tdag(z)=-\omega^{\pm{P-\Pdag\over4}}{\brks{1}z\over z'+z}
\left(\omega^{P-\Pdag\over4}s_+(z)-\omega^{\Pdag-P\over4}s_-(z)\right)+O(1)
\quad
\text{as }z'\to\e^{\pm\i\pi}z.
\label{kinpole}
$$
From (\ref{slambda-pair}) we have
\eq$$
s_\ve(z')t(z)=f_\ve\left(-{z\over z'}\right)\lcolon s_\ve(z')t(z)\rcolon,
\qquad
s_\ve(z')\tdag(z)=f_{-\ve}\left(z\over z'\right)\lcolon s_\ve(z')\tdag(z)\rcolon.
\label{st-prods}
$$
Let $X=(x_1,\ldots,x_n)$, $Y=(y_1,\ldots,y_{n'})$ and
$$
t(X)=t(x_1)\cdots t(x_n),
\qquad
\tdag(Y)=\tdag(y_1)\cdots\tdag(y_{n'}).
$$
Let $\langle u|\in\cF_{rs}$, $|v\rangle\in\bcF_{rs}$ such that $\langle u|D^\pm_{-k}=0$, $D^\pm_k|v\rangle=0$ ($k>0$).
Then from the identities (\ref{kinpole}),~(\ref{st-prods}) we easily derive that the functions $J(X|Y)=\langle u|t(X)\tdag(Y)|v\rangle$ satisfy the kinematic pole condition~(\ref{JcO-kinpole}). It means that they define form factors of quasilocal operators according to~(\ref{ff-JRfactor}).

\subsection{Form factors from free fields}

Let us recall, that form factors of arbitrary physical operators can be found as soon as the functions $J_\cO$ are fixed. Assuming
$$
J_\cO(X|Y)=J_{rs}(X|Y)\equiv{}_{rs}\langle1|t(X)\tdag(Y)1\rangle_{rs},
$$
we have~\cite{Jimbo:2000ff,Lashkevich:2016lzr}
\eq$$
\cO=\Phi^\kappa_{m\bm}={\phi^\kappa_{m\bm}\over G^\kappa_{m\bm}},
\label{checkPhi-def}
$$
if
\eq$$
r=1+{\kappa\over2}+{m-\bm\over4},
\qquad
s=1+{\kappa\over2}-{m-\bm\over4},
\qquad
n'-n={m+\bm\over2}=Q,
\label{rs-kappambm}
$$
where we denote $n=\#X$, $n'=\#Y$. The quantity $Q$ is the charge of the operator and it is the same for all form factors so that the triple $(r,s,Q)$ characterizes an exponential operator as well as the triple $(\kappa,m,\bm)$. The numbers $G^\kappa_{m\bm}$ are normalization constants. In the special case $\bm=-m$ they coincide with the vacuum expectation values $\langle\phi^\kappa_{m,-m}\rangle$ calculated in~\cite{Fateev:2008zz}.%
\footnote{The factors $G^\kappa_{m,-m}$ were calculated in~\cite{Fateev:2008zz} for the $\Z_N$ Ising model. They are easily continued to arbitrary $g$ and differ from those of the complex sinh\-/Gordon model by simple factors. Below we will need a combination that is the same in both theories.}
They will play very little role in our consideration, and we will not quote the explicit expression here.

Though we never use the explicit form of the $J$ functions below, let us give it for completeness:
\eq$$
J_{rs}(X|Y)
=\left(\prod_{i=1}^nx_i\prod_{j=1}^{n'}y_i^{-1}\right)^{r-s\over N}
\sum_{\{\ve_i=\pm\}\atop\{\ve'_j=\pm\}}
\omega^{{p\over4}\sum\ve_i+{\pdag\over4}\sum\ve'_j}
\left\langle\prod^n_{i=1}\lambda_{\ve_i}(x_i)
\prod^{n'}_{i=1}\lambdadag_{\ve'_i}(y_i)\right\rangle,
\label{Jrs-explicit}
$$
where
\eq$$
\left\langle\prod^n_{i=1}\lambda_{\ve_i}(x_i)
\prod^{n'}_{i=1}\lambdadag_{\ve'_i}(y_i)\right\rangle
=\prod_{1\le i<j\le n\atop\ve_i+\ve_j=0}f_{\ve_i}\left(x_j\over x_i\right)
\prod_{1\le i<j\le n'\atop\ve'_i+\ve'_j=0}f_{\ve_i}\left(y_j\over y_i\right)
\prod_{\substack{1\le i\le n\\1\le j\le n'\\\ve_i+\ve'_j=0}}f_{-\ve_i}\left(-{y_j\over x_i}\right).
\label{lambda-prod-explicit}
$$

Now let us turn to the problem of description of descendant operators~(\ref{descendant-def}). Though up to now it is not possible to find form factors of each operator of the form~(\ref{descendant-def}), it is possible to establish a set of solutions to the form factor equations that correspond to each family~$(\cH^\kappa_{m\bm})_{\Delta k,L}$. The basic idea is that the $J$ functions for operators in~$(\cH^\kappa_{m\bm})_{\Delta k,L}/(\cH^\kappa_{m\bm})_{\Delta k,L-1}$ are obtained as matrix elements of the form $\langle u|t(X)\tdag(Y)|v\rangle$, where $\langle u|\in(\cF_{rs})_k$, $|v\rangle\in(\bcF_{rs})_{\bar k}$, so that $\Delta k=k-\bar k$, $L=k+\bar k$. Generally, matrix elements do not provide form factors since the corresponding $F$ functions obtained by means of~(\ref{ff-JRfactor}) do not satisfy the kinematic pole equation~(\ref{ffax-kinematic}). But, as we discussed at the end of the last subsection, this equation is satisfied subject to $\langle u|D^\pm_{-k}=0$, $D^\pm_k|v\rangle=0$ ($k>0$). Such vectors are generated from the vacuums by the elements of the Heisenberg algebra that commute with $D^\pm_k$.

Explicitly, let $\cA=\bigoplus^\infty_{k=0}\cA_k$ be a graded commutative algebra with the generators $a_{-k},b_{-k}$, $k\in\Z_{>0}$. The grading $\cA_k$ is defined according to the rule $\deg a_{-k}=\deg b_{-k}=k$. Consider two representations of this algebra in the Heisenberg algebra~(\ref{ddcommut}). Slightly abusing the notation, let
\eq$$
\Aligned{
a_{-k}
&={\alpha^-_k-\alpha^+_k\over A^{(1)}_k},
\qquad
&\bar a_{-k}
&={\alpha^-_{-k}-\alpha^+_{-k}\over A^{(1)}_k},
\\
b_{-k}
&={\beta^-_k-\beta^+_k\over A^{(2)}_k},
&\bar b_{-k}
&={\beta^-_{-k}-\beta^+_{-k}\over A^{(2)}_k}.
}\label{piab-def}
$$
The representatives without bar act on the bra states, while those with bar act on the ket ones. Let $h\in\cA$. Then the states
\eq$$
{}_{rs}\langle h|={}_{rs}\langle1|h,
\qquad
|h\rangle_{rs}=\bh|1\rangle_{rs}
\label{h-states}
$$
satisfy the equations
\eq$$
{}_{rs}\langle h|D_{-k}=0,
\qquad
D_k|h\rangle_{rs}=0.
\label{Dk-cond}
$$
Moreover, they provide all solutions to these equations.

Let $h,h'\in\cA$. We claim then that the functions 
\eq$$
J^{h\bh'}_{rs}(X|Y)={}_{rs}\langle h|t(X)\tdag(Y)|h'\rangle_{rs}
\label{Jhbh-def}
$$
define according to~(\ref{ff-JRfactor}) form factors of a descendant of the exponential operator $\Phi^\kappa_{m\bm}$, which will be denoted symbolically as $h\bh'\Phi^\kappa_{m\bm}$. If $h\in\cA_k$ and $h'\in\cA_{\bar k}$, the corresponding operator belongs to the subspace $(\cH^\kappa_{m\bm})_{k-\bar k,k+\bar k}$. It means that its spin and scaling dimension are defined by~(\ref{sd-descendant}).

Note that an operator of the form $h\Phi^\kappa_{m\bm}$ (which means that $h'=1$) is a right (`chiral') descendant, i.e.\ a descendant of level~$(k,0)$. Since its total level~$L$ coincides with its external spin $\Delta k$, the last is well defined and no operators of lower dimensions admix except for special resonant values of the parameters. The same is valid for left (`antichiral') descendants $\bh'\Phi^\kappa_{m\bm}$, with the only difference that their total levels~$L$ coincide with~$-\Delta k$.

Let us make a remark about identities for the functions $J^{h\bh'}_{rs}(X|Y)$. First, it is trivial that
\eq$$
{}_{r+N,s+N}\langle h|t(X)\tdag(Y)|h'\rangle_{r+N,s+N}
=(-1)^Q\times{}_{rs}\langle h|t(X)\tdag(Y)|h'\rangle_{rs}.
\label{r+N-s+N-idntity}
$$
It means that the shift $r\to r+N$, $s\to s+N$ (or, equivalently, $\kappa\to\kappa+2N$) only multiplies the operator by $(-1)^Q$:
\eq$$
h\bh'\Phi^{\kappa+2N}_{m\bm}(x)=(-1)^{(m+\bm)/2}h\bh'\Phi^\kappa_{m\bm}(x).
\label{r+N-s+N-symmetry}
$$
The second identity was proved in~\cite{Lashkevich:2016lzr} and called there the reflection identity. Namely, there are linear maps $\rho_{rs}:\cA\to\cA$, which commute with the grading, such that $\rho_{rs}(1)=1$ and
\eq$$
{}_{rs}\langle h|t(X)\tdag(Y)|\rho_{rs}(h')\rangle_{rs}
=C_{Q,rs}\times{}_{-s-N+1,-r-N+1}\langle\rho_{rs}(h)|t(X)\tdag(Y)|h'\rangle_{-s-N+1,-r-N+1},
\label{reflection}
$$
where
\eq$$
C_{Q,rs}=\prod^Q_{i=1}{\brks{s+{Q\over2}-i}\over\brks{r-{Q\over2}-1+i}}.
\label{reflection-const}
$$
In the operator language
\eq$$
h\,\overline{\rho_{rs}(h')}\,\Phi^\kappa_{m\bm}=C_{Q,rs}\rho_{rs}(h)\bh'\Phi^{-\kappa-2N-2}_{m\bm}.
\label{reflection-op}
$$
The third identity reads
\eq$$
{}_{r+N,s}\langle h|t(X)\tdag(Y)|h'\rangle_{r+N,s}
=(-1)^n\prod^n_{i=1}x_i\prod^{n'}_{i=1}y_i^{-1}{}_{rs}\langle h|t(X)\tdag(Y)|h'\rangle_{rs}.
\label{r+N-idntity}
$$
It means that the operator $h\bh'\Phi^{\kappa+N}_{m+N,\bm-N}$ differs from the operator $h\bh'\Phi^\kappa_{m\bm}$ by a simple factor in form factors. More generally, for every operator $\cO$ there exists $\cO^{(1)}$ such that
\eq$$
F_{\cO^{(1)}}(\theta_1,\ldots,\theta_n|\theta'_1,\ldots,\theta'_{n'})
=(-1)^n\exp\left(\sum^n_{i=1}\theta_i-\sum^{n'}_{i=1}\theta'_i\right)
F_\cO(\theta_1,\ldots,\theta_n|\theta'_1,\ldots,\theta'_{n'}),
\label{chain-property}
$$
so that the charge and spin of the new operator are $Q^{(1)}=Q$, $s^{(1)}=s-Q$. Continuing the procedure, we obtain an infinite chain of operators
\eq$$
\ldots,\ \cO^{(-2)},\ \cO^{(-1)},\ \cO^{(0)}\equiv\cO,\ \cO^{(1)},\ \cO^{(2)},\ldots
\label{cO-chain}
$$
This property is related with a symmetry of the Zamolodchikov\--Faddeev algebra in the case of diagonal $S$ matrix for a particle\-/antiparticle pair. Due to the absence of such symmetry in the $\Z_N$ Ising model, no more than one operator of each such chain can be consistent with the reduction from the complex sine\-/Gordon model to the $\Z_N$ Ising model.

\subsection{Computation of matrix elements}
\label{subsec-comput-matel}

In the last subsection we presented an explicit expression for the $J$ functions corresponding to exponential operators~$\Phi^\kappa_{m\bm}$. Here we will show that there is a straightforward way to obtain the $J$ functions (and, due to (\ref{ff-JRfactor}), form factors) for any operator $h\bh'\Phi^\kappa_{m\bm}$, once we know the explicit form of~$h,h'$ in terms of the generators $a_{-k},b_{-k}$. This construction is based on the following three sets of commutation relations. First, the commutation relations between the representatives $a_{-k},b_{-k}$ and the exponents $\lambda_\ve(z),\lambdadag_\ve(z)$ together with the action on the ket vacuums read
\eq$$
\Gathered{
\Aligned{{}
[a_{-k},\lambda_\ve(z)]
&=(-\ve)^{k+1}z^k\lambda_\ve(z),
\quad
&[b_{-k},\lambda_\ve(z)]
&=(-\ve)^kz^k\lambda_\ve(z),
\\
[a_{-k},\lambdadag_\ve(z)]
&=(-\ve)^{k+1}z^k\lambdadag_\ve(z),
&[b_{-k},\lambdadag_\ve(z)]
&=-(-\ve)^kz^k\lambdadag_\ve(z),
}\\
a_{-k}|1\rangle_{rs}=b_{-k}|1\rangle_{rs}=0.
}\label{ab-lambda-commut}
$$
These commutation relations are sufficient to compute all functions $J^h_{rs}$ corresponding to the right descendants. Similarly, the commutation relations between the representatives $\bar a_{-k},\bar b_{-k}$ and the exponents $\lambda_\ve(z),\lambdadag_\ve(z)$ together with the action on the bra vacuums read
\eq$$
\Gathered{
\Aligned{{}
[\lambda_\ve(z),\bar a_{-k}]
&=\ve^{k+1}z^{-k}\lambda_\ve(z),
\quad
&[\lambda_\ve(z),\bar b_{-k}]
&=\ve^kz^{-k}\lambda_\ve(z),
\\
[\lambdadag_\ve(z),\bar a_{-k}]
&=\ve^{k+1}z^{-k}\lambdadag_\ve(z),
&[\lambdadag_\ve(z),\bar b_{-k}]
&=-\ve^kz^{-k}\lambdadag_\ve(z),
}\\
{}_{rs}\langle1|\bar a_{-k}={}_{rs}\langle1|\bar b_{-k}=0.
}\label{barab-lambda-commut}
$$
These commutation relations are sufficient to compute all functions $J^\bh_{rs}$ corresponding to the left descendants. Nevertheless, both (\ref{ab-lambda-commut}) and (\ref{barab-lambda-commut}) are not sufficient to compute general $J^{h\bh}_{rs}$ functions. To do it we need the third set of relations:
\eq$$
\Gathered{
\Aligned{{}
[a_{-k},\bar a_{-l}]
&=-(1+(-1)^k)k\delta_{kl}/A^{(1)}_k,
\\
[b_{-k},\bar b_{-l}]
&=-(1-(-1)^k)k\delta_{kl}/A^{(2)}_k,
}\\
[a_{-k},\bar b_{-l}]=[b_{-k},\bar a_{-l}]=0,
}\label{ab-commut}
$$

As a result, we obtain the following expression
\eq$$
J^{h\bh'}_{rs}(X|Y)
=\left(\prod_{i=1}^nx_i\prod_{j=1}^{n'}y_i^{-1}\right)^{r-s\over N}
\sum_{\{\ve_i=\pm\}\atop\{\ve'_j=\pm\}}
\omega^{{p\over4}\sum\ve_i+{\pdag\over4}\sum\ve'_j}
P^{h\bh'}_{E|E'}(X|Y)
\left\langle\prod^n_{i=1}\lambda_{\ve_i}(x_i)
\prod^{n'}_{i=1}\lambdadag_{\ve'_i}(y_i)\right\rangle.
\label{Jhbhrs-explicit}
$$
Here $E=\{\ve_1,\ldots,\ve_n\}$, $E'=\{\ve'_1,\ldots,\ve'_{n'}\}$, and the Laurent polynomials $P^{h\bh'}_{E|E'}(X|Y)$, which linearly depend on the algebraic elements $h,h'$, are defined by the following inductive procedure. First, $P^1_{E|E'}(X|Y)=1$. Second, the `elementary' polynomials are defined explicitly
\eq$$
\Aligned{
P^{a_{-k}}_{E|E'}(X|Y)
&=\sum^n_{i=1}(-\ve_i)^{k+1}x_i^k+\sum^{n'}_{i=1}(-\ve'_i)^{k+1}y_i^k,
\\
P^{b_{-k}}_{E|E'}(X|Y)
&=\sum^n_{i=1}(-\ve_i)^kx_i^k-\sum^{n'}_{i=1}(-\ve'_i)^ky_i^k,
\\
P^{\bar a_{-k}}_{E|E'}(X|Y)
&=\sum^n_{i=1}\ve_i^{k+1}x_i^{-k}+\sum^{n'}_{i=1}\ve_i^{\prime k+1}y_i^{-k},
\\
P^{\bar b_{-k}}_{E|E'}(X|Y)
&=\sum^n_{i=1}\ve_i^kx_i^{-k}-\sum^{n'}_{i=1}\ve_i^{\prime k}y_i^{-k}.
}\label{P-elementary}
$$
Now let
$$
h=\prod_ka_{-k}^{p_k}b_{-k}^{q_k},
\qquad
h'=\prod_ka_{-k}^{p'_k}b_{-k}^{q'_k}
$$
with finite numbers of nonunit factors, i.e. the sequences $p_k\ge0$ etc.\ stabilizing at zero. Besides, let
$$
h_{k|}=\Cases{h|_{p_k\to p_k-1},&p_k>0\\0,&p_k=0,}
\qquad
h_{|k}=\Cases{h|_{q_k\to q_k-1},&q_k>0\\0,&q_k=0,}
$$
and similarly for $h'_{k|},h'_{|k}$. Then the inductive relations are
\eq$$
\Aligned{
P^{ha_{-k}\bh'}_{E|E'}(X|Y)
&=P^{h\bh'}_{E|E'}(X|Y)-kp'_k{1+(-1)^k\over A^{(1)}_k}P^{h\bh'_{k|}}_{E|E'}(X|Y),
\\
P^{hb_{-k}\bh'}_{E|E'}(X|Y)
&=P^{h\bh'}_{E|E'}(X|Y)-kq'_k{1-(-1)^k\over A^{(2)}_k}P^{h\bh'_{|k}}_{E|E'}(X|Y),
\\
P^{h\bh'\bar a_{-k}}_{E|E'}(X|Y)
&=P^{h\bh'}_{E|E'}(X|Y)-kp_k{1+(-1)^k\over A^{(1)}_k}P^{h_{k|}\bh'}_{E|E'}(X|Y),
\\
P^{h\bh'\bar b_{-k}}_{E|E'}(X|Y)
&=P^{h\bh'}_{E|E'}(X|Y)-kq_k{1-(-1)^k\over A^{(2)}_k}P^{h_{|k}\bh'}_{E|E'}(X|Y).
}\label{P-induction}
$$
Note that the second terms in these relations are the result of `gluing' the right and the left chiralities. They vanish in the case of chiral descendants.

Another important remark is that the action of the elements $a_{1-2k}$, $b_{-2k}$ on the operators is trivial. It is easy to see that the corresponding `elementary' polynomial are $E$ and $E'$ independent. Denote
\eq$$
\iota_{-k}=\Cases{a_{-k},&k\in2\Z+1;\\b_{-k},&k\in2\Z,}
\qquad
\iota_k=\bar\iota_{-k}=\Cases{\bar a_{-k},&k\in2\Z+1;\\\bar b_{-k},&k\in2\Z,}
\label{iota-def}
$$
for $k>0$. Hence, we have
$$
F_{\iota_{-k}\cO}(\theta_1,\ldots,\theta_{n+n'})_{\alpha_1\ldots\alpha_{n+n'}}
=\sum^{n+n'}_{i=1}(-1)^{(k+1)\delta_{\alpha_k\bar1}}\e^{k\theta_i}F_{\cO}(\theta_1,\ldots,\theta_{n+n'})_{\alpha_1\ldots\alpha_{n+n'}}
$$
for $k\ne0$. But the prefactor in the r.h.s.\ is easily recognized as the action of the integral of motion~$I_k$ on the operator~$\cO$:
\eq$$
[\cO(x),I_k]=\cN_k\iota_{-k}\cO(x),
\label{cOIk-commut}
$$
where $\cN_k$ is an appropriate normalization factor. For the cases $k=\pm1$ the currents $I_{\pm1}$ may be chosen to coincide with the components $P_\pm$ of the momentum, and the normalization factors read $\cN_1=-\cN_{-1}=-M/2$ in terms of the physical mass~$M$ of the particles.

The existence of explicit expressions for form factors makes, on the first glance, the whole free field representation unnecessary. The only point, where it proved its efficiency in this section was a simple proof of the kinematic pole. Nevertheless, as we will see in the next section, in its framework we introduce new objects --- screening currents and screening operators, --- which make it possible to create interesting families of descendant operators and prove important theorems about them.

\section{Screening currents and screening operators}
\label{sec-screening}

In~\cite{Lashkevich:2013mca,Lashkevich:2013yja,Lashkevich:2014rua,Lashkevich:2014qna} we constructed screening currents and screening operators in the case of the sinh\-/Gordon model. We have shown that these operators make it possible to prove important identities, which are analogous to the resonance identities in the conformal perturbation theory, to find form factors of conserved currents, and to study reductions in the sine\-/Gordon theory. In the present paper we start to follow this program for the complex sinh\-/Gordon model. In this section we construct screening currents and screening operators and discuss their properties. We construct these objects in analogy to the algebraic structures that appear in the deformed parafermion theory~\cite{Jimbo:2000ff}.

\subsection{Screening currents}

Let us introduce screening currents in terms of the Heisenberg algebra (\ref{ddcommut}),~(\ref{dP-commut}). Later we will use them to construct screening operators, i.e.\ special integrals of these currents that commute with $t(z),\tdag(z)$. There are two types of screening currents. The type~I currents are defined as%
\footnote{It may seem strange that we associate $S$ to $\tomega$ and $\tS$ to $\omega$, but we do it so that the notation stressed the analogy with our previous work on the sinh\-/Gordon model. There we used the symbol $q$ for $\tomega$ and $\tq$ for~$\omega$.}
\eq$$
\Aligned{
S(z)
&=\e^{2\ud+2\uddag}\omega^{{P-\Pdag\over8}+{\gamma^+-\gamma^-\over2}}
\lcolon\exp\sum_{k\ne0}\left({\alpha^-_k-\alpha^+_k\over\tbrks{k}}
-{\beta^-_k-\beta^+_k\over\tbrcs{k}}\right){z^{-k}\over k}\rcolon,
\\
\Sdag(z)
&=\e^{2\ud+2\uddag}\omega^{{\Pdag-P\over8}+{\gamma^--\gamma^+\over2}}
\lcolon\exp\sum_{k\ne0}\left({\alpha^-_k-\alpha^+_k\over\tbrks{k}}
+{\beta^-_k-\beta^+_k\over\tbrcs{k}}\right){z^{-k}\over k}\rcolon.
}\label{Scur-def}
$$
The type~II currents look even simpler:
\eq$$
\Aligned{
\tS(z)
&=\e^{-2\ud-(2N+2)\uddag}z^{\gamma^+-\gamma^-}
\lcolon\exp\sum_{k\ne0}{\alpha^-_k-\alpha^+_k+\beta^-_k-\beta^+_k\over\brks{k}}
{z^{-k}\over k}\rcolon,
\\
\tSdag(z)
&=\e^{-(2N+2)\ud-2\uddag}z^{\gamma^--\gamma^+}
\lcolon\exp\sum_{k\ne0}{\alpha^-_k-\alpha^+_k-\beta^-_k+\beta^+_k\over\brks{k}}
{z^{-k}\over k}\rcolon.
}\label{tScur-def}
$$
Nevertheless, their behavior is more complicated because of the factor~$z^{\pm(\gamma^+-\gamma^-)}$. We will see that this results in a kind of mode number shifts.

To save the space on the number of equations, we will omit some definitions and identities from now on. Namely, for any operator $X$ the operator $\Xdag$ is defined according to the rule
\eq$$
X\to\Xdag
\quad\Leftrightarrow\quad
P\leftrightarrow\Pdag,\
\ud\leftrightarrow\uddag,\
\gamma^\ve\rightarrow-\gamma^\ve,\
\beta^\ve_k\rightarrow-\beta^\ve_k.
\label{Xdag-def}
$$
Furthermore, for any identity that contains some operators without and with a dagger there exists another identity, which is obtained by the substitution
\eq$$
X\leftrightarrow X^\dagger:\
S\leftrightarrow\Sdag,\
\tS\leftrightarrow\tSdag,\
t\leftrightarrow\tdag,\
P\leftrightarrow\Pdag,\
\ud\leftrightarrow\uddag,\
\text{etc.}
\label{X-Xdag-equivalence}
$$
for every operator~$X$ in the identity.

The algebra of screening currents is defined in terms of operator products. These are given by
\eq$$
\Aligned{
S(z')S(z)
&=\left(1-{z\over z'}\right)\lcolon S(z')S(z)\rcolon=-{z\over z'}S(z)S(z'),
\\
\Sdag(z')S(z)
&=\left(1+{z\over z'}\right)\lcolon\Sdag(z')S(z)\rcolon={z\over z'}S(z)\Sdag(z')
}\label{SS-OP}
$$
for the type~I operators. These relations were obtained straightforwardly by using the commutation relations between the oscillators. There is a similar relation between the type~II screening currents as well:
\eq$$
\Aligned{
\tS(z')\tS(z)
&=\left(1-{z\over z'}\right)\lcolon\tS(z')\tS(z)\rcolon=-{z\over z'}\tS(z)\tS(z'),
\\
\tSdag(z')\tS(z)
&=\left(1+{z\over z'}\right)\lcolon\tSdag(z')\tS(z)\rcolon={z\over z'}\tS(z)\tSdag(z')
}\label{tStS-OP}
$$
As for relation between the currents of different types, we obtain
\eq$$
\Aligned{
S(z')\tS(z)
&=-\i\left(1+\i{z\over z'}\right)^{-1}\lcolon S(z')\tS(z)\rcolon={z'\over z}\tS(z)S(z'),
\\
S(z')\tSdag(z)
&=\i\left(1-\i{z\over z'}\right)^{-1}\lcolon S(z')\tSdag(z)\rcolon={z'\over z}\tSdag(z)S(z')
}\label{StS-OP}
$$
These operator products define completely the commutation relations in the algebra of modes of screening operators. 

For any operator $X(z)$ constructed of the oscillators let us define its modes
\eq$$
X_k=\oint{dz\over2\pi\i}z^{k-1}X(z).
\label{Xk-def}
$$
We will assume that the integral is taken over a closed contour that encloses all poles that come from the operator products of $X(z)$ with the operators written to the right of it, and does not enclose any pole that comes from the operator products with the operators written to the left.

Then from the operator products (\ref{SS-OP}), (\ref{tStS-OP}) we easily obtain
\eq$$
\Aligned{
S_kS_l
&=-S_{l+1}S_{k-1},
\quad
&S_k\Sdag_l
&=\Sdag_{l+1}S_{k-1},
\\
\tS_k\tS_l
&=-\tS_{l+1}\tS_{k-1},
\quad
&\tS_k\tSdag_l
&=\tSdag_{l+1}\tS_{k-1}.
}\label{SS-commut}
$$
Since the operator products (\ref{StS-OP}) contain poles, the corresponding commutation relations have terms that come from their residues:
\eq$$
\Aligned{
S_k\tS_l-\tS_{l-1}S_{k+1}
&=\i^{-1-k}\epsilon_{k+l}\omega^{-{k+l\over2}+{P-\Pdag\over8}},
\\
S_k\tSdag_l-\tSdag_{l-1}S_{k+1}
&=\i^{k+1}\epsilondag_{k+l}\omega^{{k+l\over2}+{P-\Pdag\over8}}.
}\label{StS-commut}
$$
Here the r.h.s.\ contain the modes of the operators $\epsilon(z),\epsilondag(z)$ defined according to
\eq$$
\epsilon(z)
=\e^{-2N\uddag}z^{\gamma^+-\gamma^-}
\exp\sum_{k\ne0}{(1-(-1)^k)(\alpha^-_k-\alpha^+_k)+(1+(-1)^k)(\beta^-_k-\beta^+_k)\over\brks{2k}}
{z^{-k}\over k},
\label{epsilon-def}
$$
and the rule~(\ref{Xdag-def}). It is easy to check that
\eq$$
\Aligned{
\epsilon(z')S(z)
&=\lcolon\epsilon(z')S(z)\rcolon=\i S(z)\epsilon(z'),
\quad
&\epsilon(z')\Sdag(z)
&=\lcolon\epsilon(z')\Sdag(z)=-\i\Sdag(z)\epsilon(z'),
\\
\epsilon(z')\tS(z)
&=\lcolon\epsilon(z')\tS(z)\rcolon=\tS(z)\epsilon(z'),
\quad
&\epsilon(z')\tSdag(z)
&=\lcolon\epsilon(z')\tSdag(z)=\tSdag(z)\epsilon(z').
}\label{epsS-OP}
$$
Hence,
\eq$$
\epsilon_kS_l=\i S_l\epsilon_k,
\qquad
\epsilon_k\Sdag_l=-\i\Sdag_l\epsilon_k,
\qquad
\epsilon_k\tS_l=\tS_l\epsilon_k,
\qquad
\epsilon_k\tSdag_l=\tSdag_l\epsilon_k.
\label{epsS-commut}
$$

\subsection{Screening operators and null vectors}

Now we construct the operators that commute with the currents $t(z),\tdag(z)$. We will see that the existence of these operators result in identities between descendant operators at special values of the parameters $\kappa,m,\bm$, namely
for integer values of $\kappa$, $(\kappa-m)/2$ or, equivalently, for integer values of~$r+Q/2,s+Q/2$.

Consider the operator products of the screening currents with the exponential operators $\lambda,\lambdadag$. For the type~I currents we have
\eq$$
\Aligned{
S(z')\lambda_+(z)
&=\omega^{-1/4}\lcolon S(z')\lambda_+(z)\rcolon=\omega^{-1/2}\lambda_+(z)S(z'),
\\
S(z')\lambda_-(z)
&=\omega^{-1/4}{z'+\tomega^{-1/2}z\over z'-\tomega^{1/2}z}\lcolon S(z')\lambda_-(z)\rcolon
=\omega^{1/2}\lambda_-(z)S(z'),
\\
S(z')\lambdadag_+(z)
&=\omega^{1/4}{z'+\tomega^{1/2}z\over z'-\tomega^{-1/2}z}\lcolon S(z')\lambdadag_+(z)\rcolon
=\omega^{-1/2}\lambdadag_+(z)S(z'),
\\
S(z')\lambdadag_-(z)
&=\omega^{1/4}\lcolon S(z')\lambdadag_-(z)\rcolon=\omega^{1/2}\lambdadag_-(z)S(z').
}\label{Slambda-OP}
$$
The corresponding identities for $\Sdag$ are obtained by means of the substitution~(\ref{X-Xdag-equivalence}) as it was mentioned in the last subsection.

What we need is the commutation relations of $S_k,\Sdag_k$ with $t(z)$, $\tdag(z)$. Nonzero commutators come from the singularities in the operator products. We have
\eq$$
\Aligned{{}
[S_k,t(z)]
&=\i\tbrcs{1}\,z^k\sigma(z)\omega^{-{P+\Pdag-2\over8}}\tomega^{k/2},
\\
[S_k,\tdag(z)]
&=-\i\tbrcs{1}\,z^k\sigmadag(z)\omega^{P+\Pdag-2\over8}\tomega^{-k/2},
}\label{St-commut}
$$
where $\sigma(z),\sigmadag(z)$ are defined by
\eq$$
\sigma(z)
=\e^{2\ud+2\uddag}z^{P-\Pdag\over2N}\omega^{\gamma^++\gamma^-\over2}
\lcolon\exp\sum_{k\ne0}\left({\tomega^{k/2}\alpha^-_k-\tomega^{-k/2}\alpha^+_k\over\tbrks{k}}
+{\tomega^{k/2}\beta^-_k+\tomega^{-k/2}\beta^+_k\over\tbrcs{k}}\right){z^{-k}\over k}\rcolon.
\label{sigma-def}
$$
We see that the r.h.s.\ of identities (\ref{St-commut}) are nonzero for any values of $r,s$. Thus, we need to combine them with the corresponding identities for $\Sdag_k$ to cancel these terms. Consider the combinations
\eq$$
S^\pm_k=S_k\pm(-1)^{k-1}\Sdag_k.
\label{Spm-def}
$$
Later we will need their commutation relations with each other:
\eq$$
S^\ve_kS^{\ve'}_l+S^{\ve'}_{l+1}S^\ve_{k-1}=0,
\label{SpmSpm-commut}
$$
and with~$\tS,\tSdag$:
\eq$$
\Aligned{
S^\pm_k\tS_l-\tS_{l-1}S^\mp_{k+1}
&=\i^{-1-k}\epsilon_{k+l}\left(\omega^{{P-\Pdag\over8}-{k+l\over2}}\pm\omega^{-{P-\Pdag\over8}+{k+l\over2}}\right),
\\
S^\pm_k\tSdag_l-\tSdag_{l-1}S^\mp_{k+1}
&=\i^{k+1}\epsilondag_{k+l}\left(\omega^{{P-\Pdag\over8}+{k+l\over2}}\pm\omega^{-{P-\Pdag\over8}-{k+l\over2}}\right).
}\label{Spm-tS-commut}
$$
For the commutators of these combinations with $t,\tdag$ we have
\eq$$
\Aligned{{}
[S^\pm_k,t(z)]
&=\i^{1-k}\tbrcs{1}z^k\sigma(z)
\left(\omega^{-{P+\Pdag-2\over8}-{k\over2}}\pm\omega^{{P+\Pdag-2\over8}+{k\over2}}\right),
\\
[S^\pm_k,\tdag(z)]
&=\i^{k-1}\tbrcs{1}z^k\sigmadag(z)
\left(\omega^{{P+\Pdag-2\over8}+{k\over2}}\pm\omega^{-{P+\Pdag-2\over8}-{k\over2}}\right).
}\label{Spm-t-commut}
$$
Here the r.h.s.\ contain sums of two terms, which can be canceled at special values of~$r,s$. Define the operators
\eq$$
\Aligned{
\Sigma^+
&=S^+_k:\ \bcF_{rs}\to\bcF_{r-1,s-1},
\quad
&&{\textstyle k=-{\kappa\over2}=1-{r+s\over2}};
\\
\Sigma^-
&=S^-_k:\ \bcF_{rs}\to\bcF_{r-1,s-1},
\quad
&&{\textstyle k=-{\kappa+N\over2}=1-{r+s+N\over2}},
}\label{Sigma-def}
$$
These operators commute with $t(z)$, $\tdag(z)$ for integer values of~$k$:
\eq$$
\Aligned{
&[\Sigma^+,t(z)]|_{\bcF_{rs}}=[\Sigma^+,\tdag(z)]|_{\bcF_{rs}}=0,
&&\text{if $r+s\in2\Z$,}
\\
&[\Sigma^-,t(z)]|_{\bcF_{rs}}=[\Sigma^-,\tdag(z)]|_{\bcF_{rs}}=0,
&&\text{if $r+s+N\in2\Z$.}
}\label{Sigma-t-commut}
$$
These values of $r,s$ correspond to even values of $\kappa$ or $\kappa+N$. Later we will use $\Sigma^+$ in the $\cH^0_{00}$ space of operators, where the conserved currents $T_k$ appear.

Now turn to the type~II screening currents. We have
\eq$$
\Aligned{
\tS(z')\lambda_\pm(z)
&={z^{\prime1/2}\over z'-\omega^{\mp1/2}z}\lcolon\tS(z')\lambda_\pm(z)\rcolon
=-\omega^{\pm1/2}z^{-1}\lambda_\pm(z)\tS(z'),
\\
\tS(z')\lambdadag_\pm(z)
&={z'+\omega^{\pm1/2}z\over z^{\prime1/2}}\lcolon\tS(z')\lambdadag_\pm(z)\rcolon
=\omega^{\pm1/2}z\lambdadag_\pm(z)\tS(z').
}\label{tSlambda-OP}
$$
These operator products contain poles for both $\lambda_+$ and $\lambda_-$ so that the r.h.s.\ of the corresponding commutation relations with $t,\tdag$ have the desired form of sums at once. The result is written in terms of anticommutators:
\eq$$
\{\tS_k,t(z)\}=z^k\tsigma(z)\brcs{P-2k+1},
\qquad
\{\tS_k,\tdag(z)\}=0,
\label{tSt-commut}
$$
where
\Align$$
\tsigma(z)
&=\e^{-2\ud-(2N+2)\uddag} z^{\gamma^+-\gamma^-+{P-\Pdag\over2N}-{1\over2}}\omega^{\gamma^++\gamma^-\over2}
\notag
\\*
&\quad\times
\lcolon\exp\sum_{k\ne0}{\omega^{k/2}(\alpha^-_k+\beta^-_k)-\omega^{-k/2}(\alpha^+_k+\beta^+_k)
\over\brks{k}}{z^{-k}\over k}\rcolon.
\label{tsigma-def}
$$
The operator $\tsigmadag$ appears in the corresponding commutator of~$\tSdag_k$. Looking at the r.h.s.\ of~(\ref{tSt-commut}) we see that it vanishes when the brace vanishes, in particular, when $r=k$. Define
\eq$$
\Aligned{
\tSigma
&=\tS_r:\ \bcF_{rs}\to\bcF_{r+1,s+N+1},
\quad
&&r+Q/2\in\Z;
\\
\tSigmadag
&=\tSdag_s:\ \bcF_{rs}\to\bcF_{r+N+1,s+1},
\quad
&&s+Q/2\in\Z.
}\label{tSigma-def}
$$
The requirement that $r+Q/2$ or $s+Q/2$ is an integer is not quite obvious. Let us explain it. The power at the exponent in $\tS_r$ contains, beside of the explicit factor $z^{r-1}$, the factor $z^{\gamma^+-\gamma^-}$ (see the definition~(\ref{tScur-def})). It is easy to see that for any function $f$
\eq$$
\langle\omega^{a\gamma^++b\gamma^-}f(\gamma^+-\gamma^-)\omega^{a'\gamma^++b'\gamma^-}\rangle
=f\left(a+b-a'-b'\over2\right)\langle\omega^{(a+a')\gamma^++(b+b')\gamma^-}\rangle.
\label{gammagamma-eigenvalue}
$$
It is related to the hidden parameter of the vacuum, mentioned above after equation~(\ref{gamma-normal}). Since in each matrix element $a+b$ and $a'+b'$ are integers and $a+b+a'+b'=-Q$, the eigenvalue of $\gamma^+-\gamma^-$ is always an integer, if $Q$ is even, and a half\-/integer, if $Q$ is odd. Since the total power of $z$ under the integral must be integer, we have $r+Q/2\in\Z$.

We have
\eq$$
\Aligned{
\{\tSigma,t(z)\}|_{\bcF_{rs}}
&=\{\tSigma,\tdag(z)\}|_{\bcF_{rs}}=0,
\quad
&&r+Q/2\in\Z;
\\
\{\tSigmadag,t(z)\}|_{\bcF_{rs}}
&=\{\tSigmadag,\tdag(z)\}|_{\bcF_{rs}}=0,
\quad
&&s+Q/2\in\Z.
\label{tSigma-commut}
}
$$
From the commutation relations (\ref{SS-commut}), (\ref{SpmSpm-commut}), (\ref{Spm-tS-commut}) we obtain
\eq$$
\Gathered{
(\Sigma^\pm)^2=0,
\qquad
\tSigma^2=0,
\qquad
(\tSigmadag)^2=0,
\\
\tSigma\tSigmadag=\e^{2N(\ud-\uddag)}\tSigmadag\tSigma\e^{-2N(\ud-\uddag)},
\\
\Sigma^+\tSigma=\tSigma\Sigma^-,
\qquad
\Sigma^-\tSigma=\tSigma\e^{2N(\ud+\uddag)}\Sigma^+\e^{-2N(\ud+\uddag)},
\\
\Sigma^+\tSigmadag=\tSigmadag\Sigma^-,
\qquad
\Sigma^-\tSigmadag=\tSigmadag\e^{2N(\ud+\uddag)}\Sigma^+\e^{-2N(\ud+\uddag)},
}\label{Sigma-relations}
$$
while from (\ref{epsS-commut}) we have
\eq$$
\Gathered{
\epsilon_k\Sigma^+=\i\Sigma^-\epsilon_k,
\qquad
\epsilon_k\Sigma^-=\i\e^{-4N\ud}\Sigma^+\e^{4N\ud}\epsilon_k,
\\
\epsilon_k\tSigma=\e^{-2N\ud}\tSigma\e^{2N\ud}\epsilon_k,
\qquad
\epsilon_k\tSigmadag=\tSigmadag\epsilon_k.
}\label{Sigma-epsilon-relations}
$$

Now let us discuss applications of screening operators to form factors. Let $r+Q/2,s+Q/2\in\Z$. From (\ref{tSigma-commut}) we immediately obtain
\eq$$
{}_{r+1,s+N+1}\langle h|\tS_rt(X)\tdag(Y)|h'\rangle_{rs}
=(-1)^Q\times{}_{r+1,s+N+1}\langle h|t(X)\tdag(Y)\tS_r|h'\rangle_{rs}.
\label{matel-tS-example}
$$
This produces an identity for form factor. Consider the right action of $\tS_r$ on the vector ${}_{r+1,s-N+1}\langle h|$. Due to equation~(\ref{gammagamma-eigenvalue}) it depends on the charge $Q$, but we need a $Q$ independent action on the element~$h$. Thus, we define the algebraic element $s_k(h)$ according to
\eq$$
{}_{rs}\langle\tilde s_{-k}(h)|={}_{r+1,s+N+1}\langle h|\tS_{k-Q/2}.
\label{tsk(h)-def}
$$
Then the identity (\ref{matel-tS-example}) reads
\eq$$
{}_{rs}\langle\tilde s_{-r-Q/2}(h)|t(X)\tdag(Y)|h'\rangle_{rs}
=(-1)^Q\times{}_{r+1,s+N+1}\langle h|t(X)\tdag(Y)|\tilde s_{r-Q/2}(h')\rangle_{r+1,s+N+1}
\label{matel-tS-rewritten}
$$
On the operator side it means
\eq$$
\tilde s_{-1-(\kappa+m)/2}(h)\bh'\Phi^\kappa_{m\bm}(x)
=(-1)^{m+\bm\over2}h\,\overline{\tilde s_{1+(\kappa-\bm)/2}(h')}\,\Phi^{\kappa+N+2}_{m-N,\bm+N}(x).
\label{Phi-tSid-example}
$$
An interesting example is the case $h=h'=1$ and $\kappa+m,\kappa-\bm\ge0$. In this case
\eq$$
\tilde s_{-1-(\kappa+m)/2}(1)\Phi^\kappa_{m\bm}(x)=0,
\label{Phi-null}
$$
which corresponds to a null vector in conformal field theory.

Now suppose $r+s-2=\kappa=-2k$ be even. Then the $\Sigma^+$ operator is defined and
\eq$$
{}_{r-1,s-1}\langle h|S^+_kt(X)\tdag(Y)|h'\rangle_{rs}
={}_{r-1,s-1}\langle h|t(X)\tdag(Y)S^+_k|h'\rangle_{rs}.
\label{matel-Splus-example}
$$
With the definition
\eq$$
{}_{rs}\langle s_{-k}(h)|={}_{r-1,s-1}\langle h|S^+_k
\label{sk(h)-def}
$$
we have an operator identity
\eq$$
s_{\kappa/2}(h)\bh'\Phi^\kappa_{m\bm}(x)=h\,\overline{s_{-\kappa/2}(h')}\,\Phi^{\kappa-2}_{m\bm}(x),
\label{Phi-Sid-example}
$$
which corresponds to a set of resonance identities in the order $|\kappa|/2$ of the conformal perturbation theory in~$\chi\bar\chi$.

\subsection{Inverse screening currents}

Now introduce some more objects: inverse type~I screening currents and their modes. The inverse screening currents $\Sinv(z),\Sdaginv(z)$ are defined by formal inversion of the currents~(\ref{Scur-def}) under the normal order sign (do not forget about the rule~(\ref{Xdag-def})):
\eq$$
\Sinv(z)
=\e^{-2(\ud+\uddag)}\omega^{{\Pdag-P\over8}+{\gamma^--\gamma^+\over2}}
\lcolon\exp\sum_{k\ne0}\left({\alpha^+_k-\alpha^-_k\over\tbrks{k}}
-{\beta^+_k-\beta^-_k\over\tbrcs{k}}\right){z^{-k}\over k}\rcolon.
\label{Sinv-cur-def}
$$
There modes will also be labeled by the $-1$ superscript:
\eq$$
\Sinv_k=\int{dz\over2\pi\i}\,z^{k-1}\Sinv(z),
\qquad
\Sdaginv_k=\int{dz\over2\pi\i}\,z^{k-1}\Sdaginv(z).
\label{Sinv-k-def}
$$
We will need their commutation relations between themselves:
\eq$$
\Sinv_k\Sinv_l=-\Sinv_{l+1}\Sinv_{k-1},
\qquad
\Sinv_k\Sdaginv_l=\Sdaginv_{l+1}S_{k-1},
\label{Sinv-Sinv-commut}
$$
and with modes of the screening currents:
\eq$$
\Aligned{
\Sinv_kS_l+S_{l-1}\Sinv_{k+1}
&=\delta_{k+l,0},
&\Sinv_k\Sdag_l-\Sdag_{l-1}\Sinv_{k+1}
&=(-1)^l\eta_{k+l},
\\
\Sinv_k\tS_l-\tS_{l+1}\Sinv_{l-1}
&=0,
&\Sinv_k\tSdag_l-\tSdag_{l+1}\Sinv_{l-1}
&=0,
}\label{Sinv-S-commut}
$$
where $\eta_k$ and $\etadag_k$ are modes of the currents $\eta(z)$, $\etadag(z)$:
\Align$$
\eta(z)
&=\omega^{{\Pdag-P\over4}+\gamma^--\gamma^+}
\notag
\\*
&\quad\times
\exp\sum_{k\ne0}{z^{-k}\over k}\left(
  {(1-(-1)^k)(\alpha^+_k-\alpha^-_k)\over\tbrks{k}}
  -{(1+(-1)^k)(\beta^+_k-\beta^-_k)\over\tbrcs{k}}
\right){z^{-k}\over k}.
\label{eta-def}
$$
We will also need their commutators with $\epsilon_k$, $\epsilondag_k$:
\eq$$
\epsilon_k\Sinv_l=-\i\Sinv_l\epsilon_k,
\qquad
\epsilon_k\Sdaginv_l=\i\Sdaginv_l\epsilon_k.
\label{epsSinv-commut}
$$
The commutation relations with $t(z),\tdag(z)$ read
\eq$$
\Aligned{{}
[\Sinv_k,t(z)]
&=(-1)^k\omega^{1/4}\tbrcs{1}\,\tomega^{-{k-1\over2}}
\lcolon\eta(-\tomega^{-1/2}z)\ttau(z)\rcolon z^{k+{P-\Pdag\over2N}}\omega^{-{P+\Pdag\over8}},
\\
[\Sdaginv_k,t(z)]
&=(-1)^k\omega^{-1/4}\tbrcs{1}\,\tomega^{k-1\over2}
\lcolon\etadag(-\tomega^{1/2}z)\ttau(z)\rcolon z^{k+{P-\Pdag\over2N}}\omega^{P+\Pdag\over8},
}\label{Sinv-t-commut}
$$
where
\eq$$
\ttau(z)
=\e^{-2(\ud+\uddag)}\omega^{{\gamma^++\gamma^-\over2}}
\lcolon\exp\sum_{k\ne0}\left({\tomega^{k/2}\alpha^+_k-\tomega^{-k/2}\alpha^-_k\over\tbrks{k}}
  +{\tomega^{k/2}\beta^+_k+\tomega^{-k/2}\beta^-_k\over\tbrcs{k}}\right)
{z^{-k}\over k}.
\label{ttau-def}
$$
We also have
\eq$$
\Sinv_l[\Sinv_k,t(z)]=-[\Sinv_{k+1},t(z)]\Sinv_{l-1},
\qquad
\Sinv_l[\Sdaginv_k,t(z)]=[\Sdaginv_{k+1},t(z)]\Sinv_{l-1},
\label{Sinv-t-double-commut}
$$

The inverse screening operators make it possible to define elements $\ft_{-\vec k|-\vec l}$ according to
\eq$$
{}_{rs}\langle\ft_{-k_1,\ldots,-k_m|-l_1,\ldots,-l_n}|
\omega^{\left({1\over4}(r-s)+{1\over2}(\gamma^+-\gamma^-)\right)(n-m)}
={}_{r+m+n,s+m+n}\langle1|\Sinv_{k_1}\cdots\Sinv_{k_m}\Sdaginv_{l_1}\cdots\Sdaginv_{l_n}.
\label{tinvstates-def}
$$
These elements can also be defined in terms of the ket vectors:
\eq$$
\omega^{\left({1\over4}(r-s)+{1\over2}(\gamma^+-\gamma^-)\right)(n-m)}
|\ft_{-k_1,\ldots,-k_m|-l_1,\ldots,-l_n}\rangle_{rs}
=\Sinv_{l_n}\cdots\Sinv_{l_1}\Sdaginv_{k_m}\cdots\Sdaginv_{k_1}|1\rangle_{r-m-n,s-m-n}.
\label{tinvstates-ket-def}
$$
Notice that the screening operators with and without daggers exchanged their places. For simplicity we will write
$$
\ft_{-\vec k}=\ft_{-\vec k|\varnothing},
\qquad
\ftdag_{-\vec k}=\ft_{\varnothing|-\vec k}.
$$
These simplest elements are easily obtained from the series
\eq$$
\sum^\infty_{k=1}\ft_{-k}z^k
=\exp\sum^\infty_{k=1}B_k(a_{-k}+b_{-k})z^k,
\qquad
B_k={\brks{k}\over2\i^kk}.
\label{tinvstates-series}
$$
Explicitly, the first few of them look like
\subeq{\label{tinvstates-examples}
\Align$$
\ft_{-1}
&=B_1(a_{-1}+b_{-1}),
\label{tinvstates-examples-1}
\\
\ft_{-2}
&=B_2(a_{-2}+b_{-2})+{B_1^2\over2}(a_{-1}+b_{-1})^2,
\label{tinvstates-examples-2}
\\
\ft_{-3}
&=B_3(a_{-3}+b_{-3})+B_1B_2(a_{-1}+b_{-1})(a_{-2}+b_{-2})+{B_1^3\over6}(a_{-1}+b_{-1})^3.
\label{tinvstates-examples-3}
$$}
The elements $\ftdag_{-k}$ are obtained from $\ft_{-k}$ by the substitution $b_{-k}\to-b_{-k}$.

\section{Conserved currents}
\label{sec-conscur}

Above we discussed two methods to construct descendant operators in terms of free fields. The first method was based on the algebra~$\cA$ and its representations. This method seems to be general, at least, at generic values of the parameters. The second approach is based on the screening currents, and it is not quite clear, how general it is. In this section we demonstrate the second method on a rather simple example.

\subsection{The conservation laws in the form factor language}

An important problem of the bootstrap form factor approach is to find the relation between the operators defined in terms of their form factors (in our approach the operators $h\bh\Phi^\kappa_{m\bm}$) and those defined in terms of the fields $\vartheta,\varphi$ of the dual complex sinh\-/Gordon theory. In general, finding this relation is a rather sophisticated problem. In this section we would like to concentrate on a class of physically interesting descendant operators, namely, the conserved currents (or densities of local integrals of motion) $T_{\pm k}(x)$, $\Theta_{\pm(k-2)}(x)$ ($k\ge2$), as we discussed in subsection~\ref{subsec-intr-intpert}. The currents $T_k$ are right chiral descendants of the unit operator $\phi^0_{00}$, while $\Theta_{k-2}$ can be chosen in different ways, depending of what perturbation theory we are starting from. We will choose the sine\-/Liouville theory (for the complex sinh\-/Gordon model) or $\Z_N$ parafermion theory (for the $\Z_N$ Ising model) for a nonperturbed conformal theory, and the operator $\phi^2_{00}\sim\chi\bar\chi$ for the perturbation. In this picture the operators $\Theta_{k-2}$ are right descendants of the perturbation operator $\phi^2_{00}$ due to the resonance phenomenon. Similarly, $T_{-k}$ and $\Theta_{2-k}$ are left descendants of $1$ and $\phi^2_{00}$ correspondingly.

Let us rewrite the continuity equations (\ref{continuity}) in terms of form factors 
\eq$$
\Aligned{
\langle\vac|\d_-T_k|\theta_1,\ldots,\theta_{2n}\rangle_{\alpha_1\ldots\alpha_{2n}}
&=\langle\vac|\d_+\Theta_{k-2}|\theta_1,\ldots,\theta_{2n}\rangle_{\alpha_1\ldots\alpha_{2n}},
\\
\langle\vac|\d_+T_{-k}|\theta_1,\ldots,\theta_{2n}\rangle_{\alpha_1\ldots\alpha_{2n}}
&=\langle\vac|\d_-\Theta_{2-k}|\theta_1,\ldots,\theta_{2n}\rangle_{\alpha_1\ldots\alpha_{2n}}
}\label{continuity-ff}
$$
for $k\ge2$. Since the space\-/time derivatives are generated by the momentum components, they may be substituted by their eigenvalues:
\Multline$$
\langle\vac|\d_\pm\cO(0)|\theta_1,\ldots,\theta_{2n}\rangle_{\alpha_1\ldots\alpha_{2n}}=
\langle\vac|[\cO(0),-\i P_\pm]|\theta_1,\ldots,\theta_{2n}\rangle_{\alpha_1\ldots\alpha_{2n}}=
\\
=\pm{\i M\over2}\sum^{2n}_{i=1}\e^{\pm\theta_i}\times
\langle\vac|\cO(0)|\theta_1,\ldots,\theta_{2n}\rangle_{\alpha_1\ldots\alpha_{2n}}.
\label{dbd-exp}
$$
But the expressions $\sum^{2n}_{i=1}\e^{\pm\theta_i}=\sum^n_{i=1}x_i^{\pm1}+\sum^n_{i=1}y_i^{\pm1}$ are obtained by the action of the elements $a_{-1}$ for upper signs and $\bar a_{-1}$ for lower signs. Thus, we have
\eq$$
\bar a_{-1}T_k=-a_{-1}\Theta_{k-2},
\qquad
a_{-1}T_{-k}=-\bar a_{-1}\Theta_{2-k},
\qquad
k\ge2.
\label{continuity-Aalg}
$$
In other words, our aim is to find for each $k\ge2$ two pairs $h,h'\in\cA_k$ and $g,g'\in\cA_{k-2}$ such that
\eq$$
\Aligned{
{}_{11}\langle h|t(X)\tdag(Y)|a_{-1}\rangle_{11}
&=-{}_{22}\langle a_{-1}|t(X)\tdag(Y)|g\rangle_{22},
\\
{}_{11}\langle a_{-1}|t(X)\tdag(Y)|h'\rangle_{11}
&=-{}_{22}\langle g'|t(X)\tdag(Y)|a_{-1}\rangle_{22}.
\label{continuity-conjecture}
}
$$
Then $T_k=h1$, $\Theta_{k-2}=g\Phi^2_{00}$, $T_{-k}=\bar h'1$, $\Theta_{2-k}=\bar g'\Phi^2_{00}$. In the case $k=2$ we know that $g=g'=\langle\Theta_0\rangle$ and, due to the known vacuum expectation value of the operator $\phi^2_{00}$ and the coupling constant $\lambda$ in terms of the mass~$M$, it is possible to find $T_2,T_{-2},\Theta_0$ so that they were related to the energy\-/momentum tensor according to the standard convention:
\eq$$
T_2=-2\pi T_{++},
\qquad
T_{-2}=-2\pi T_{--},
\qquad
\Theta_0=2\pi T_{+-}.
\label{T2-20-norm}
$$
From~\cite{Fateev:1993av,Fateev:2008zz} we have%
\footnote{The values of $\lambda$ and $\langle\phi^2_{00}\rangle$ were found for the $\Z_N$ Ising model, but their product can be analytically continued to the complex sinh\-/Gordon model.}
\eq$$
\langle\Theta_0\rangle=\pi{N\over N+2}\lambda\langle\phi^2_{00}\rangle
={\pi M^2\over4\sin{2\pi\over N}}.
\label{Theta0-VEV}
$$
Due to the continuity equations, it fixes the normalization of $T_{\pm2}$ as well.

\subsection{Explicit construction of conserved currents}

Return to the elements $\ft_{-k},\ftdag_{-k}$ defined in~(\ref{tinvstates-series}). It is easy to prove the identities:
\eq$$
\Aligned{
{}_{11}\langle\ft_{-k}|t(X)\tdag(Y)|a_{-1}\rangle_{11}
&={}_{-N-1,-N-1}\langle a_{-1}\ft^{-1}_{2-k}|t(X)\tdag(Y)|1\rangle_{-N-1,-N-1},
\\
{}_{11}\langle\ftdag_{-k}|t(X)\tdag(Y)|a_{-1}\rangle_{11}
&={}_{-N-1,-N-1}\langle a_{-1}\ftdag_{2-k}|t(X)\tdag(Y)|1\rangle_{-N-1,-N-1}.
}\label{continuity-ft}
$$
To do it, first, notice that
\eq$$
\Aligned{
{}_{rs}\langle a_{-1}|\brcs{1}
&={}_{r+N+2,s+N+2}\langle1|\tS_1\tSdag_0,
\\
\brcs{1}|a_{-1}\rangle_{rs}
&=\tS_0\tSdag_{-1}|1\rangle_{r-N-2,s-N-2}.
}\label{am1-screenings}
$$
Then we have
\Align*$$
{}_{11}\langle\ft_{-k}|t(X)\tdag(Y)|a_{-1}\rangle_{11}
&={}_{22}\langle1|\{1\}_\omega^{-1}\Sinv_kt(X)\tdag(Y)\tS_0\tSdag_{-1}|1\rangle_{-N-1,-N-1}
\\
&={}_{22}\langle1|\{1\}_\omega^{-1}\Sinv_kt(X)\tdag(Y)\tSigma\e^{2N\uddag}\tSigmadag\e^{-2N\uddag}|1\rangle_{-N-1,-N-1}
$$
Both $\tSigma$ and $\e^{2N\uddag}\tSigmadag\e^{-2N\uddag}$ anticommute with $t(z)$, $\tdag(z)$. Hence, we may push them to the left:
\Align*$$
{}_{11}\langle\ft_{-k}|t(X)\tdag(Y)|a_{-1}\rangle_{11}
&={}_{22}\langle1|\{1\}_\omega^{-1}\Sinv_k\tSigma\e^{2N\uddag}\tSigmadag\e^{-2N\uddag}t(X)\tdag(Y)|1\rangle_{-N-1,-N-1}
\\
&={}_{22}\langle1|\{1\}_\omega^{-1}\Sinv_k\tS_0\tSdag_{-1}t(X)\tdag(Y)|1\rangle_{-N-1,-N-1}
$$
Then, by using the commutation relations (\ref{Sinv-S-commut}) and again (\ref{am1-screenings}), we obtain
\Align*$$
{}_{11}\langle\ft_{-k}|t(X)\tdag(Y)|a_{-1}\rangle_{11}
&={}_{22}\langle1|\{1\}_\omega^{-1}\tS_1\tSdag_0\Sinv_{k-2}t(X)\tdag(Y)|1\rangle_{-N-1,-N-1}
\\
&={}_{-N,-N}\langle a_{-1}|\Sinv_{k-2}t(X)\tdag(Y)|1\rangle_{-N-1,-N-1}
\\
&={}_{-N-1,-N-1}\langle a_{-1}\ft_{2-k}|t(X)\tdag(Y)|1\rangle_{-N-1,-N-1},
$$
which proves the first equation of~(\ref{continuity-ft}). The second one is proved similarly.

Equations~(\ref{continuity-ft}) are very similar to equations~(\ref{continuity-conjecture}) with one difference. In the r.h.s.\ they contain states in the modules with $r=s=-N-1$ instead of $r=s=2$. Nevertheless, due to the reflection property~(\ref{reflection-op}) the r.h.s.\ of~(\ref{continuity-ft}) corresponds to descendants of the same operator~$\Phi^{-N-1}_{00}=\Phi^2_{00}$.

So we have two candidates for the continuity equation, corresponding to that of $T_k$, $k\ge2$. Now we claim that they are physically the same. Let
\eq$$
\fT^\pm_{-k}={1\over2}(\ft_{-k}\pm(-1)^k\ftdag_{-k})
\label{Tk-Thetak-V}
$$
In what follows we will also use the notation
\eq$$
(\Sinv_k)^\pm={1\over2}(\Sinv_k\pm(-1)^k\Sdaginv_k),
\label{Sinvpm-def}
$$
so that
$$
{}_{rr}\langle\fT^\pm_{-k}|={}_{r+1,r+1}\langle1|(\Sinv_k)^\pm,
\qquad
|\fT^\pm_{-k}\rangle_{rr}=\pm(-1)^k(\Sinv_{-k})^\pm|1\rangle_{r-1,r-1}.
$$

Let us show that the combination $\fT^-_{-k}$ corresponds to a linear combination of operators of the form $\sum^{k-1}_{l=1}[I_k,(\text{something})]$. Indeed, it can be written as
$$
\fT^-_{-k}=\sum^\infty_{K,L=0}
\sum_{\substack{\{k_i\}^K_{i=1},\{l_j\}^L_{j=1}\\\sum k_i+\sum l_j=k}}A_{\{k_i\}\{l_j\}}
\prod^K_{i=1}a_{-k_i}\prod^L_{j=1}b_{-l_j}.
$$
First, consider the case of odd $k$. Then, the terms with even $L$ only enter the sum. Hence, if $\sum l_j$ is even, the sum $\sum k_i$ is odd and, hence, at least one $k_i$ is odd. On the contrary, if $\sum l_j$ is odd, and at least one $l_j$ is even. Second, consider the case of even $k$. The terms with odd $L$ only enter the sum. If $\sum l_j$ is odd, the sum $\sum k_i$ is odd too, and there is at least one odd $k_i$. If $\sum l_j$ is even, at least one even $l_j$ must exist. Hence, in both cases at least one odd $k_i$ or even $l_j$ enters each term. But $a_{-k_i}$ with odd $k_i$ as well as $b_{-l_j}$ with even $l_j$ corresponds to a commutator with an integral of motion according to (\ref{iota-def}),~(\ref{cOIk-commut}).

Hence, the element $\ftdag_{-k}$ do not produce any new independent conserved currents in comparison with $\ft_{-k}$. We will use their combination $\fT^+_{-k}$ as the `shortest' element corresponding to a conserved current. Thus, we identify
\eq$$
T_k=G_k\fT^+_{-k}\Phi^0_{00},
\qquad
\Theta_{k-2}=-G_k\fT^+_{2-k}\Phi^{-2N-4}_{00},
\qquad
k\ge2,
\label{Tk-identification}
$$
where
\eq$$
G_k={\i^k\pi M^k\over4\sin{2\pi\over N}}.
\label{GTTheta-def}
$$
This normalization factor is chosen in consistency with~(\ref{Theta0-VEV}) for~$k=2$, but for $k>2$ it is arbitrary and only provides the correct dimension.

In a similar way we obtain the identification for $T_{-k}$, $\Theta_{2-k}$:
\eq$$
T_{-k}=-G_k\bar\fT^+_{-k}\Phi^{-2}_{00},
\qquad
\Theta_{2-k}=G_k\bar\fT^+_{2-k}\Phi^{2N+2}_{00}.
\label{Tmk-identification}
$$
The operator $\Phi^{-2}_{00}$ is identified with $\Phi^0_{00}$ and the operator $\Phi^{2N+2}_{00}$ with $\Phi^2_{00}$ via the reflection and $2N$ (quasi)periodicity:
$$
\Phi^{-2}_{00}=\Phi^{-2N}_{00}=\Phi^0_{00},
\qquad
\Phi^{2N+2}_{00}=\Phi^2_{00}.
$$
Note that though the $2N$ periodicity directly identifies descendant operators as well as the exponential ones, the reflection demands applying the reflection map $\rho_{rs}$, which is proved to exist, but generally not known explicitly.

Now we have to discuss the property expressed in equations (\ref{r+N-idntity})--(\ref{cO-chain}). It is evident that the whole chains of the operators $T_k^{(p)}$, $\Theta_k^{(p)}$ ($p\in\Z$) satisfy the same continuity equations:
\eq$$
\Aligned{
\d_-T^{(p)}_k
&=\d_+\Theta^{(p)}_{k-2},
\\
\d_+T^{(p)}_{-k}
&=\d_-\Theta^{(p)}_{-k+2}.
}\label{p-continuity}
$$
But since non\-/diagonal matrix elements of the integrals of motion $I_k$ vanish, while the diagonal ones are $p$ independent, there is no difference between these operators with different values of $p$: we may substitute $T^{(p)}_k$, $\Theta^{(p)}_k$ from any values of $p$ into~(\ref{Ik-in-currents}) instead of $T_k$, $\Theta_k$. It means that there exist operators $\Xi^{(p)}_k(x)$ such that
\eq$$
\Aligned{
T^{(p)}_k-T_k
&=\d_+\Xi^{(p)}_{k-1},
\quad
&\Theta^{(p)}_{k-2}-\Theta_{k-2}
&=\d_-\Xi^{(p)}_{k-1},
\\
T^{(p)}_{-k}-T_{-k}
&=\d_-\Xi^{(p)}_{1-k},
\quad
&\Theta^{(p)}_{2-k}-\Theta_{2-k}
&=\d_+\Xi^{(p)}_{1-k}.
}\label{Xi(p)-def}
$$
We will use this fact while constructing products of currents in the next section for the products $T_kT_{-l}$. By using the definition~(\ref{TT-def}), from~(\ref{Xi(p)-def}) it is easy to prove that $T^{(p)}_kT^{(p')}_{-l}-T_kT_{-l}$ is a combinations of space\-/time derivatives for any values of $p,p'$. Indeed,
\Multline*$$
(T^{(p)}_k(x+\epsilon)-T_k(x+\epsilon))T^{(p')}_{-l}(x)
-(\Theta^{(p)}_{k-2}(x+\epsilon)-\Theta_{k-2}(x+\epsilon))\Theta^{(p')}_{2-l}(x)
\\
=\d_+\Xi^{(p)}_{k-1}(x+\epsilon)T^{(p')}_{-l}(x)-\d_-\Xi^{(p)}_{k-1}(x+\epsilon)\Theta^{(p')}_{2-l}(x)
\\
=\d_\mu(\cdots)^\mu-\Xi^{(p)}_{k-1}(x+\epsilon)(\d_+T^{(p')}_{-l}(x)-\d_-\Theta^{(p')}_{2-l}(x))
=\d_\mu(\cdots)^\mu
$$
and the same for the $T^{(p')}_{-l}-T_{-l}$. Since we defined $T_kT_{-l}$ up to space\-/time derivatives, it is sufficient find form factors for any operator $T^{(p)}_kT^{(p')}_{-l}$ or even for any their linear combination. The question about consistency with the reduction to the $\Z_N$ Ising model is more subtle, and we will not discuss it here. Anyway, it does not affect diagonal matrix elements like (\ref{TT-ff}), and we may ignore it in the context of studying the perturbations~(\ref{ZN-pert}).

\section{Form factors for the products \texorpdfstring{$T_kT_{-l}$}{Tk T-k}}
\label{sec-TT}

\subsection{The main conjecture}

We have shown that each current $T_k$ ($k\ge2$) corresponds, up to a factor, to the bra vector ${}_{11}\langle\fT_{-k}|$, while $T_{-k}$ corresponds to the ket vector $|\fT_{-k}\rangle_{00}$. We would like to unite them into one matrix element ${}_{11}\langle\fT_{-k}|\cdots|\fT_{-l}\rangle_{00}$, which, as we expect, will describe the product $T_kT_{-l}$. The problem is that the operator denoted by dots must shift the zero mode. For the sinh\-/Gordon model~\cite{Lashkevich:2014qna} we used the possibility to define representative for both $T_k$, $T_{-l}$ in the same Fock module. Here we have no such possibility. Instead, we may insert~$\tS_0=\Sigma$ or $\tSdag_0=\tSigmadag$. We may conjecture that the matrix element
$$
{}_{11}\langle\fT^+_{-k}|t(X)\tdag(Y)\tS_0|\fT^+_{-l}\rangle_{0,-N}
$$
corresponds to the product $T_kT^{(-1)}_{-l}$, which, as we discussed at the end of the last subsection only differs from $T_kT_{-l}$ by space\-/time derivatives. To verify our conjecture, let us check its consistency with the asymptotic cluster factorization principle~\cite{Delfino:1996nf}. Namely, split each of the sets $X$ and $Y$ into two disjoint nonempty parts:
$$
X=X_R\sqcup X_L,
\qquad Y=Y_R\sqcup Y_L,
\qquad
0<\#X_R=\#Y_R<n.
$$
Let $\Lambda$ be a real variable. Consider the matrix element
$$
{}_{11}\langle\fT^+_{-k}|t(\e^\Lambda X_R)t(X_L)\tdag(\e^\Lambda Y_R)\tdag(Y_L)\tS_0|\fT^+_{-l}\rangle_{0,-N}
$$
in for large positive $\Lambda$. Since $\tS_0$ is a screening operator $\tSigma$, we may rewrite it as
$$
{}_{11}\langle\fT^+_{-k}|t(\e^\Lambda X_R)\tdag(\e^\Lambda Y_R)\tS_0t(X_L)\tdag(Y_L)|\fT^+_{-l}\rangle_{0,-N}
$$
In the limit $\Lambda\to+\infty$ the main contribution comes from the vacuum intermediate state near $\tS_0$ and, hence, the matrix element factorizes:
\Multline*$$
{}_{11}\langle\fT^+_{-k}|t(\e^\Lambda X_R)\tdag(\e^\Lambda Y_R)|1\rangle_{11}
\times{}_{0,-N}\langle1|t(X_L)\tdag(Y_L)|\fT^+_{-l}\rangle_{0,-N}
\\
=\e^{k\Lambda}\times{}_{11}\langle\fT^+_{-k}|t(X_R)\tdag(Y_R)|1\rangle_{11}
\times{}_{0,-N}\langle1|t(X_L)\tdag(Y_L)|\fT^+_{-l}\rangle_{0,-N}.
$$
Since $R(\Lambda)\Rdag(\Lambda)\to1$ as $\Lambda\to\pm\infty$, it means that each form factor splits into the product of a form factor of $T_k$ and that of $T^{(-1)}_{-l}$ in this limit in consistency with the mentioned principle.

The insertion of $\tSdag_0$ corresponds to $T_kT^{(1)}_{-l}$. To minimize the number of space\-/time derivative terms it is convenient to take the sum of the two types of matrix elements. Namely, define the functions
\eq$$
J_{T_kT_{-l}}(X|Y)
={G_kG_l\over2}
\times{}_{11}\langle\fT^+_{-k}|t(X)\tdag(Y)(\tS_0\epsilondag_0+\tSdag_0\epsilon_0)|\fT^+_{-l}\rangle_{-N,-N}.
\label{Tkl-def}
$$
We will think that these functions define the form factors of the operator $T_kT_{-l}$ (modulo space\-/time derivatives). The $\epsilon_0$, $\epsilondag_0$ operators are inserted in (\ref{Tkl-def}) for the sake of shifts of zero modes solely.

As a check, let us calculate the vacuum expectation value of the spinless operators $T_kT_{-k}$. By definition, we have
\Align*$$
\langle T_kT_{-k}\rangle=J_{T_kT_{-k}}(\varnothing|\varnothing)
&=(-1)^k{\i G_k^2\over2}
\bigl({}_{22}\langle1|(\Sinv_k)^+\tS_0(\Sinv_{-k})^-|1\rangle_{-1,-1-N}
\\*
&\quad
-{}_{22}\langle1|(\Sinv_k)^+\tSdag_0(\Sinv_{-k})^-|1\rangle_{-1-N,-1}\bigr).
$$
By using (\ref{Sinv-S-commut}), (\ref{epsSinv-commut}) we obtain
\Align*$$
\langle T_kT_{-k}\rangle
&=(-1)^{k-1}{\i G_k^2\over2}
\bigl({}_{22}\langle1|(\Sinv_{2-k})^+\tS_0(\Sinv_{k-2})^-|1\rangle_{-1,-1-N}
\\
&\quad-{}_{22}\langle1|(\Sinv_{2-k})^+\tSdag_0(\Sinv_{k-2})^-|1\rangle_{-1-N,-1}\bigr).
$$
We immediately conclude that $\langle T_kT_{-k}\rangle=0$ for $k\ge3$. Thus continue for $k=2$. Since
$$
{}_{22}\langle1|(\Sinv_0)^+={}_{11}\langle1|,
\qquad
\i\tS_0(\Sinv_0)^-|1\rangle_{-1,-1-N}=-\i\tSdag_0(\Sinv_0)^-|1\rangle_{-1-N,-1}=|1\rangle_{11}
$$
for $Q=0$, we obtain
\eq$$
\langle T_2T_{-2}\rangle=-G_2^2=-\langle\Theta_0\rangle^2
\label{TbT-VEV}
$$
in consistency with~\cite{Zamolodchikov:2004ce}.

Let us study the functions~$J_{T_kT_{-l}}$ in more detail. First of all, from (\ref{Sinv-S-commut}), (\ref{epsSinv-commut}) we have
\Multline*$$
\tS_0\epsilondag_0|\fT^+_{-l}\rangle_{-N,-N}
=(-1)^l\tS_0\epsilondag_0(\Sinv_{-l})^+|1\rangle_{-1-N,-1-N}
=(-1)^l\i\tS_0(\Sinv_{-l})^-\epsilondag_0|1\rangle_{-1-N,-1-N}
\\*
=(-1)^l\i(\Sinv_{1-l})^+\tS_{-1}\epsilondag_0|1\rangle_{-1-N,-1-N}.
$$
Similarly,
$$
\tSdag_0\epsilondag_0|\fT^+_{-l}\rangle_{-N,-N}
=(-1)^{l-1}\i(\Sinv_{1-l})^+\tSdag_{-1}\epsilon_0|1\rangle_{-1-N,-1-N}.
$$
Here we have taken into account that on ket vectors $\Sinv_{-k}$ produces $\ftdag_{-k}$, while $\Sdaginv_{-k}$ produces $\ft_{-k}$. Then we easily obtain
$$
(\tS_{-1}\epsilondag_0-\tSdag_{-1}\epsilon_0)|1\rangle_{-1-N,-1-N}=2\i B_1|b_{-1}\rangle_{00}.
$$
By commuting $\Sinv_{1-l}$, $\Sdaginv_{1-l}$ with $\bar b_{-1}$ we get a more explicit expression:
\eq$$
J_{T_kT_{-l}}(X|Y)
=G_kG_l\times{}_{11}\langle\fT^+_{-k}|\,t(X)\tdag(Y)\,|B_1b_{-1}\fT^+_{1-l}-\fT^+_{-l}\rangle_{11}.
\label{Tkl-explicit}
$$
This expression is appropriate for calculating form factors with the help of the procedure described in subsection~\ref{subsec-comput-matel}. It is convenient for small values of~$k,l$, but for large values it becomes too complicated. The initial form~(\ref{Tkl-def}) turns out to be much more practical for finding expressions for form factors for general values of~$k,l$.

\subsection{Computing matrix elements by pulling inverse screening operators}

To compute the matrix elements (\ref{Tkl-def}) we have to pull $(\Sinv_k)^+$ to the right and $(\Sinv_{-l})^+$ to the left and then apply them to the vacuum. First, pull $(\Sinv_k)^+$ to the right:
\Multline*$$
J_{T_kT_{-l}}(X|Y)
=(-1)^{l-1}{G_kG_l\over2}\bigl(
{}_{22}\langle1|t(X)\tdag(Y)(\Sinv_{2-l})^+(\tS_0\epsilondag_0+\tSdag_0\epsilon_0)
(\Sinv_{k-2})^+|1\rangle_{-1-N,-1-N}
\\*
-{}_{22}\langle1|[(\Sinv_k)^+,t(X)\tdag(Y)](\Sinv_{1-l})^+
\i(\tS_{-1}\epsilondag_0-\tSdag_{-1}\epsilon_0)|1\rangle_{-1-N,-1-N}\bigr).
$$
The first term in the parentheses vanishes if $k>2$. We have to pull $(\Sinv_{2-l})^+$ there to the left. In the second term we have to expand the commutators and pull $(\Sinv_{1-l})^+$ and then $\tS_{-1},\tSdag_{-1}$ to the left. By using the commutation relations~(\ref{Sinv-t-double-commut} we obtain
\Multline*$$
J_{T_kT_{-l}}(X|Y)
=G_kG_l\bigl(
\delta_{k,2}\delta_{l,2}\delta_{n,0}
-\delta_{k,2}\times{}_{22}\langle1|t(X)\tdag(Y)|\fT^+_{2-l}\rangle_{22}
\\*
-{\delta_{l,2}\over2}\times{}_{0,-N}\langle\fT^+_{2-k}|t(X)\tdag(Y)|1\rangle_{0,-N}
-{\delta_{l,2}\over2}\times{}_{-N,0}\langle\fT^+_{2-k}|t(X)\tdag(Y)|1\rangle_{-N,0}\bigr)
+J_{[T_kT_{-l}]}(X|Y).
$$
The first two terms in parentheses correspond to the fourth and the second terms in~(\ref{TTred-def}). The sum of the third and the fourth terms is proportional to $\delta_{l,2}(\Theta^{(1)}_{k-2}+\Theta^{(-1)}_{k-2})$ and, up to space\-/time derivative terms, which we ignore, gives the third term of~(\ref{TTred-def}). The rest of the expression corresponds to the $[T_kT_{-l}]$ operator and is given by
\Multline$$
J_{[T_kT_{-l}]}(X|Y)
=(-1)^{l-1}B_1G_kG_l\times{}_{22}\langle1|\bigl([(\Sinv_k)^+,t(X)\tdag(Y)](\Sinv_{1-l})^+
\\
+(\Sinv_{2-l})^+[(\Sinv_{k-1})^+,t(X)\tdag(Y)]\bigr)|b_{-1}\rangle_{00}.
\label{J[TT]-def}
$$
More explicitly, equation~(\ref{J[TT]-def}) can be written as
\Align$$
{J_{[T_kT_{-l}]}(X|Y)\over(-1)^lB_1G_kG_l}
&=\sum^n_{i<j}\Bigl(
{}_{22}\langle1|t(\hat X_{ij})\tdag(Y)[(\Sinv_k)^+,t(x_i)][(\Sinv_{1-l})^+,t(x_j)]|b_{-1}\rangle_{00}
\notag
\\
&\qquad
-{}_{22}\langle1|t(\hat X_{ij})\tdag(Y)[(\Sinv_{2-l})^+,t(x_i)][(\Sinv_{k-1})^+,t(x_j)]|b_{-1}\rangle_{00}
\notag
\\
&\qquad
+{}_{22}\langle1|t(X)\tdag(\hat Y_{ij})[(\Sinv_k)^+,\tdag(y_i)][(\Sinv_{1-l})^+,\tdag(y_j)]|b_{-1}\rangle_{00}
\notag
\\
&\qquad
-{}_{22}\langle1|t(X)\tdag(\hat Y_{ij})[(\Sinv_{2-l})^+,\tdag(y_i)][(\Sinv_{k-1})^+,t(y_j)]|b_{-1}\rangle_{00}
\Bigr)
\notag
\\*
&\quad
+\sum^n_{i,j=1}\Bigl(
{}_{22}\langle1|t(\hat X_i)\tdag(\hat Y_j)[(\Sinv_k)^+,t(x_i)][(\Sinv_{1-l})^+,\tdag(y_j)]|b_{-1}\rangle_{00}
\notag
\\
&\qquad
-{}_{22}\langle1|[(\Sinv_{2-l})^+,t(x_i)][(\Sinv_{k-1})^+,\tdag(y_j)]|b_{-1}\rangle_{00}
\Bigr),
\label{J[TT]-explicit}
$$
where $\hat X_{ij\ldots}=X\setminus\{x_i,x_j,\ldots\}$. The expression looks a little terrifying on the first glance, but in fact it can be straightforwardly computed for arbitrary values of~$k,l$. First, we have to substitute the commutators, which are easily extracted from~(\ref{Sinv-t-commut}) and read
\eq$$
\Aligned{{}
[(\Sinv_k)^+,t(z)]|_{\bcF_{rr}}
&=-\i z^k(\tomega^{1/2}+\tomega^{-1/2})
\\
&\quad\times
\left((-1)^k\omega^{1-r\over2}\tomega^{1-k\over2}\lcolon\eta(-\tomega^{-1/2}z)\ttau(z)\rcolon
  -\omega^{r-1\over2}\tomega^{k-1\over2}\lcolon\etadag(-\tomega^{1/2}z)\ttau(z)\rcolon\right),
\\
[(\Sinv_k)^+,\tdag(z)]|_{\bcF_{rr}}
&=\i z^k(\tomega^{1/2}+\tomega^{-1/2})
\\
&\quad\times
\left((-1)^k\omega^{r-1\over2}\tomega^{k-1\over2}\lcolon\eta(-\tomega^{1/2}z)\ttaudag(z)\rcolon
  -\omega^{1-r\over2}\tomega^{1-k\over2}\lcolon\etadag(-\tomega^{-1/2}z)\ttaudag(z)\rcolon\right).  
}\label{Sinvpm-t-commut}
$$
Then, take into account the contribution of $b_{-1}$ in the ket vectors by pulling $\bar b_{-1}$ to the left with the aid of~(\ref{barab-lambda-commut}) and
\eq$$
[\ttau(z),\bar b_{-1}]=-z^{-1}\ttau(z),
\qquad
[\taudag(z),\bar b_{-1}]=z^{-1}\ttaudag(z).
\label{ttau-b-commut}
$$
At last, apply the Wick theorem to the resulting products of exponents of oscillators on the basis of the pair correlation functions
\eq$$
\Gathered{
\langle\eta(z')\eta(z)\rangle=\langle\eta(z')\etadag(z)\rangle=1,
\\
\langle\eta(z')\lambda_\pm(z)\rangle=\langle\lambda_\pm(z)\eta(z')\rangle
=\langle\eta(z')\ttau(z)\rangle=\langle\ttau(z)\eta(z')\rangle
=\omega^{1/2}{z'-\tomega^{1/2}z\over z'+\tomega^{-1/2}z},
\\
\langle\eta(z')\lambdadag_\pm(z)\rangle=\langle\lambdadag_\pm(z)\eta(z')\rangle
=\langle\eta(z')\ttaudag(z)\rangle=\langle\ttaudag(z)\eta(z')\rangle
=\omega^{-1/2}{z'-\tomega^{-1/2}z\over z'+\tomega^{1/2}z},
\\
\langle\ttau(z')\lambda_\pm(z)\rangle=\omega^{\pm1/2}\langle\lambda_\pm(z)\ttau(z')\rangle=\omega^{\pm1/4},
\\
\langle\ttau(z')\lambdadag_\pm(z)\rangle=\omega^{\pm1/2}\langle\lambdadag_\pm(z)\ttau(z')\rangle
=\omega^{\pm1/4}{\omega^{\mp1/2}z'+\omega^{\pm1/2}z\over z'+z},
\\
\langle\ttau(z')\ttau(z)\rangle=1-{z\over z'},
\qquad
\langle\ttau(z')\ttaudag(z)\rangle={(z'-\tomega z)(z'-\tomega^{-1}z)\over z'(z'+z)}.
}\label{eta-ttau-corr}
$$
In practice, the algorithm is treatable by a computer algebra system (we used \emph{Wolfram Mathematica}$^{\text{\scriptsize®}}$), or suits for numerical computations.

\subsection{Particular cases: 2-particle form factors and 2-by-2 matrix elements}

As the first example, let us give the result for the two\-/particle form factor. For $n=1$ (\ref{J[TT]-def}) reads
\Multline$$
J_{[T_kT_{-l}]}(x|y)
=(-1)^lB_1G_kG_l\bigl({}_{22}\langle1|[(\Sinv_k)^+,t(x)][(\Sinv_{1-l})^+,\tdag(y)]|b_{-1}\rangle_{00}
\\
-{}_{22}\langle1|[(\Sinv_{2-l})^+,t(x)][(\Sinv_{k-1})^+,\tdag(y)]|b_{-1}\rangle_{00}\bigr).
\label{J[TT]-2p}
$$
After a cumbersome but straightforward calculation, we obtain
\Multline$$
F_{[T_kT_{-l}]}(\theta+\delta|\theta-\delta)
={\pi^2M^{k+l}\over8\sin{2\pi\over N}}
{\e^{(k-l)\theta}\Rdag(2\delta)\sh2\delta
  \over\ch\left(\delta+{\i\pi\over N}\right)\ch\left(\delta-{\i\pi\over N}\right)}
\\*
\times\left((-1)^k\e^{-(k+l-2)\delta}H_k(\delta)H_l(\delta)
-(-1)^l\e^{(k+l-2)\delta}H_k(-\delta)H_l(-\delta)\right),
\label{F1b1-fin}
$$
where
\eq$$
H_k(\delta)=\e^{\i\pi(k-1)\over N}\ch\left(\delta+{\i\pi\over N}\right)
+\e^{-{\i\pi(k-1)\over N}}\ch\left(\delta-{\i\pi\over N}\right).
\label{Hk-def}
$$
Notice, that $F_{[T_kT_{-l}]}(\theta|\theta-\i\pi)=0$, which, in particular, means that the $[T_kT_{-l}]$ perturbations in (\ref{CShG-pert}), (\ref{ZN-pert}) do not affect the particle spectrum in the first order of the perturbation theory.

Now consider the four\-/particle form factors. A general expression for the four\-/particle form factor $F_{[T_kT_{-l}]}(\theta_1,\theta_2|\theta'_2,\theta'_1)$ is complicated, but for the diagonal case $\theta'_1=\theta_1-\i\pi$, $\theta'_2=\theta_2-\i\pi$ it simplifies. We have
\Multline$$
J_{[T_kT_{-l}]}(x_1,x_2|-x_1,-x_2)
=\pi^2M^{k+l}\brks{2k-2}\brks{2l-2}
\\*
\times{(x_1+x_2)(\omega^{1/2}x_1-\omega^{-1/2}x_2)(\omega^{-1/2}x_1-\omega^{1/2}x_2)(x_1^2x_2^{k+l}-x_1^{k+l}x_2^2)
  \over x_1^{l+2}x_2^{l+2}(x_1-x_2)}.
\label{J[TT]xxxx}
$$
Taking into account that
\eq$$
R(\theta)R(-\theta)\Rdag(\i\pi+\theta)\Rdag(\i\pi-\theta)
={\sh^2{\theta\over2}\over\sh\left({\theta\over2}+{\i\pi\over N}\right)\sh\left({\theta\over2}-{\i\pi\over N}\right)},
\label{RRRdagRdag}
$$
we obtain (\ref{TT-LNIff}) with
\eq$$
Z_k=\pi M^k{\sin{\pi(k-1)\over N}\over\sin{\pi\over N}},
\label{Zk-fin}
$$
which is easily established by computing the matrix element ${}_{11}\langle\fT^+_{-k}|t(x)\tdag(-x)|1\rangle_{11}$.

\section{Conclusion}

In this work we further develop the algebraic approach for form factors of quasilocal operators in the complex sinh\-/Gordon model. The general construction~\cite{Lashkevich:2016lzr} is based on the oscillator representation for the operators that create the particles, and on special oscillators, which generate descendant operators from the primary ones. Unlike the free boson theory, the interacting theory has inner relations between different operators, which have to be discovered to get full understanding of the space of operators. We study the structure of the space of descendant operators by introducing the algebra of screening currents. This algebra is constructed from the algebra of operators that create particles, and there is no direct analogy or guesses from the corresponding CFT. The new algebraic objects allow us to study analytically the space of form factors and to obtain new physically interesting results.

In particular, we have found null vectors in the space of descendant operators. We have also proposed closed expressions for multiparticle form factors of the conserved currents $T_k$ and $\Theta_k$ for arbitrary integer $k$ in terms of the inverse screening currents. Note that any conservation law in the form factor language is an infinite set of algebraic equations for functions of arbitrary number of variables, while the algebra of screening currents makes it possible to prove them in an elegant and simple manner.

We have proposed an expression for form factors of the composite operators $T_kT_{-l}$. These operators in the conformal limit are descendant operators with both right and left chiralities. In the conformal field theory the two chiralities are independently generated by two chiral current algebras, which include the Virasoro algebra. In the massive case there are no chiral algebras, so that two chiralities mix with each other, and this complicates studying the space of quasilocal operators. Nevertheless, the algebraic construction provides some hints to computing form factors of such operators.

With the representation for multiparticle form factors of the operators $T_kT_{-l}$ in hand, we turn to a physically interesting question of finding the scattering matrix in the current\--current integrable perturbations of the massive integrable field theories. We calculated the 2-by-2 diagonal matrix elements~(\ref{TT-LNIff}), which define the first order contribution to the scattering phases in the models~(\ref{CShG-LNIpert}). Our result confirms the conjecture about the scattering matrices of such new integrable perturbations and extends it to the case of a particle\--antiparticle pair that allow even spin integrals of motion, and to the case of Lorentz non\-/invariant perturbations.

Our results are partially applicable to the $\Z_N$ Ising models, which is a reduction of the complex sine\-/Gordon model. The representations for the currents $T_k$, $\Theta_k$ can be shown to be compatible with the reduction. The composite operators $T_kT_{-l}$ are defined modulo space\-/time derivatives and by adding such derivatives can be rendered to a reducible form. We consider a study of the reduction procedure as a natural next step.

\section*{Acknowledgments}

The authors are grateful to A.~Belavin, M.~Bershtein, A.~Litvinov and A.~Zamolodchikov for discussions. The work was supported by the Russian Science Foundation under the grant 18--12--00439. M.L.\ was able to combine scientific research with teaching owing to the support of the Simons Foundation.

\raggedright

\end{document}